\documentclass[12pt,a4,sans]{article}
\usepackage[latin1]{inputenc}
\usepackage{amssymb}
\usepackage{amsmath}
\usepackage{amsthm}
\usepackage{amsfonts}
\usepackage[pdftex]{graphicx}
\usepackage{geometry}
\usepackage{bbold}
\usepackage{braket}
\usepackage{enumerate}
\usepackage[usenames,dvipsnames]{xcolor}
\usepackage{setspace}
\usepackage[backgroundcolor=white,linecolor=black]{todonotes}
\usepackage[in]{fullpage}
\usepackage{hyperref}
\usepackage{array}
\usepackage{cancel}
\usepackage{wrapfig}
\usepackage[font=small,labelfont=bf]{caption}
\usepackage{anysize}

\graphicspath{{./images/}}

\numberwithin{equation}{section}

\marginsize{2.4cm}{2.4cm}{2cm}{2cm}

\hypersetup{
colorlinks=true,
linktoc=page,
citecolor=DarkOrchid,
linkcolor=DarkOrchid,
urlcolor=DarkOrchid} 

\makeatletter 
\renewcommand{\maketitle} 
 { \begingroup \begin{center} \large {\bf \@title}
 	\vskip 5pt \large \@author \\ \vskip 5pt \@date \end{center}
   \vskip 5pt \endgroup \setcounter{footnote}{0} }
\makeatother 

\newcommand{\comments}[1]{}
\newcommand{\la}{\langle}
\newcommand{\ra}{\rangle}
\newcommand{\tl}{\widetilde\lambda}

\newcommand{\A}{\mathcal{A}}
\newcommand{\F}{\mathcal{F}}
\newcommand{\T}{\mathcal{T}}

\newcommand{\N}{\mathcal{N}}

\newcommand{\Tr}{\text{Tr}}

\renewcommand{\b}[1]{\braket{#1}}

\renewcommand{\O}{\mathcal{O}}

\newcommand{\black}{\mathord{\parbox[c]{1em}{\includegraphics[width=0.115\textwidth]{black}}}}

\newcommand{\be}{\begin{equation}}
\newcommand{\ee}{\end{equation}}

\def\beqa{\begin{eqnarray}}
\def\eeqa{\end{eqnarray}}
\def\beq{\begin{equation}}
\def\eeq{\end{equation}}

\def\Tr{{\rm Tr}}
\def\one{\mbox{1 \kern-.59em {\rm l}}}

\def\cA{{\cal A}}  \def\cC{{\cal C}}
  \def\cF{{\cal F}}
\def\cG{{\cal G}}  
  
 \def\cN{{\cal N}} \def\cO{{\cal O}}
  \def\cR{{\cal R}}
\def\cS{{\cal S}} \def\cT{{\cal T}}

\def\uno{\mbox{1 \kern-.59em {\rm l}}}

\def\ran{\rangle}

\def\one{1\!\!1\,\,}

\def\bcomment#1{}

\def\eps{\epsilon}

\usepackage{slashed}
\usepackage{caption}

\usepackage[nottoc,notlot,notlof]{tocbibind}
\usepackage[nosort]{cite}
\usepackage{pstricks}         
\usepackage{color}
\usepackage{bbm}
\usepackage{parskip}
\usepackage{mathtools}

\graphicspath{{./Images/}}

\long\def\symbolfootnote[#1]#2{\begingroup%
\def\thefootnote{\fnsymbol{footnote}}\footnote[#1]{#2}\endgroup}

\begin{document}

\begin{flushright}
QMUL-PH-14-09
\end{flushright}

\vspace{20pt}

\begin{center}

{\Large \bf The last of the simple remainders  }

\vspace{45pt}

{\mbox {\bf Andreas Brandhuber, Brenda Penante, Gabriele Travaglini and Congkao Wen}}%
\symbolfootnote[4]{
{\tt  \{ \tt \!\!\!\!\!\! a.brandhuber, b.penante, g.travaglini, c.wen\}@qmul.ac.uk}
}

\vspace{0.5cm}

\begin{center}
{\small \em

Centre for Research in String Theory\\
School of Physics and Astronomy\\
Queen Mary University of London\\
Mile End Road, London E1 4NS, UK
}
\end{center}


\vspace{40pt}

{\bf Abstract}
\end{center}

\vspace{0.3cm}

\noindent
We compute the $n$-point two-loop form factors of the half-BPS operators ${\rm Tr}(  \phi_{AB}^n)$ in $\cN=4$ super Yang-Mills for arbitrary $n>2$ using generalised unitarity and symbols. 
These form factors are minimal in the sense that the $n^{\rm th}$ power of the scalar field in the operator requires the presence of at least $n$ on-shell legs.
Infrared divergences are shown to exponentiate as for amplitudes, reproducing the known cusp and collinear anomalous dimensions at two loops.  
We define appropriate infrared-finite remainder functions 
and compute them analytically for all $n$. 
The results obtained by using the known expressions of the integral functions involve complicated combinations of Goncharov multiple polylogarithms, but we show  that much simpler expressions can in fact be derived using the symbol of transcendental functions. 
For $n=3$ we find a very compact  remainder function expressed in terms of classical polylogarithms only. 
For arbitrary $n>3$ we are able to write  all the remainder functions in terms of a single compact building block, expressed as a sum of classical polylogarithms augmented by two  multiple polylogarithms. The decomposition of the symbol into specific components is crucial in order to single out a natural
 combination of multiple polylogarithms.
Finally, we analyse in detail the behaviour of these minimal form factors in collinear and soft limits, which deviates from the usual behaviour of amplitudes and non-minimal form factors.
 
\setcounter{page}{0}
\thispagestyle{empty}
\newpage


\setcounter{tocdepth}{4}
\hrule height 0.75pt
\tableofcontents
\vspace{0.8cm}
\hrule height 0.75pt
\vspace{1cm}

\setcounter{tocdepth}{2}
%


\pagenumbering{arabic}

\section{Introduction}

In this paper we concentrate on the calculation of two-loop form factors of half-BPS operators in $\cN=4$ supersymmetric Yang-Mills theory (SYM).
In particular we look at operators of the form $\cO_k :=\Tr (\phi_{12}^k)$, 
with $k>2$,  and their superpartners,  which can be packaged into a single superfield $\cT_k$.  
Here $\phi_{AB}=-\phi_{BA}$ denotes the three complex scalar fields of the theory, satisfying the reality condition $\bar{\phi}^{AB} =(1/2) \eps^{ABCD} \phi_{CD}$, where  $A, \ldots , D$ are $SU(4)$ {\it R}-symmetry indices.

The superfield $\cT_k$ is a generalisation of the stress-tensor multiplet $\cT_2$. For $k >2$   it is dual to massive Kaluza-Klein modes of the $AdS_5 \times S^5$ compactification of type IIB supergravity, while  for $k = 2$  it is dual to the massless graviton multiplet.

Sudakov form factors of $\cO_2$ (the lowest component of $\cT_2$) have been constructed
up to four loops \cite{vanNeerven:1985ja, Gehrmann:2011xn, Boels:2012ew}, 
while in \cite{Brandhuber:2010ad, Brandhuber:2012vm} form factors of $\cO_2$ with more than two external on-shell states were computed.
Later,   the supersymmetric form factors of $\cT_2$  were presented in \cite{Brandhuber:2011tv,Bork:2011cj} using  harmonic and Nair's on-shell superspace \cite{Nair:1988bq}, extending the results obtained for the bosonic operator $\cO_2$. 
In \cite{Bork:2010wf,Penante:2014sza} the study of form factors of $\cO_k$ and super form factors of $\cT_k$ with $k>2$ was initiated. In particular in \cite{Penante:2014sza} general expressions for MHV form factors $\cF_{\cT_k,n}$ of $\cT_k$ with $n$ external legs at tree level and one loop were presented. Such form factors have also been studied at strong coupling\cite{Maldacena:2010kp, Gao:2013dza} using AdS/CFT and integrability, exhibiting interesting similarities with amplitudes at strong coupling.\footnote{In \cite{Maldacena:2010kp} it was noted that the leading order result at strong coupling is independent of $k$. It would be interesting to find a manifestation of this fact at weak coupling.}

In this work we will focus on the special class of  form factors $\cF_{\cT_k,k}$ of $\cT_k$ which we call ``minimal", because they have the same number of on-shell legs as  scalars in $\cO_k$ (the lowest component of $\cT_k$).
In the case of Sudakov form factors $k\!=\!n\!=\!2$, and  the result has trivial kinematic dependence dictated by dimensional analysis and Lorentz invariance.
The minimal form factors $\cF_{\cT_k,k}$ are close cousins of the Sudakov form factors and, hence, it is natural to expect that their kinematic dependence will be simpler, albeit non-trivial, compared to the general case with $n>k$. Indeed, we will be able to present very compact, analytic expression for arbitrary $n=k$ written in terms of simple, universal building blocks.

Next we would like to outline briefly the strategy of our calculation. 

The first step consists in using generalised unitarity to construct the  two-loop form factors in terms of a basis of integral functions. Here we are in the fortunate situation where  all  the required integral functions are known analytically from the work of \cite{Gehrmann:1999as,Gehrmann:2000zt} in terms of classical and Goncharov polylogarithms. Such expressions are typically rather long,  but past experience \cite{DelDuca:2010zg, Goncharov:2010jf, Heslop:2011hv, Prygarin:2011gd, Brandhuber:2012vm, Dixon:2011pw,Dixon:2014voa}
suggests that for appropriate, finite  quantities,  the final result can be condensed to a much   simpler and compact form. 

Following this line of thought, and also inspired by the well-known exponentiation of infrared divergences, we will introduce finite remainder functions \cite{Anastasiou:2003kj,Bern:2005iz, Bern:2008ap, Drummond:2008aq}. These remainders are defined in terms of two important universal constants, and our calculation confirms that they coincide with the cusp anomalous dimension and collinear anomalous dimension which appear in the definition of remainders of amplitudes \cite{Bern:2008ap, Drummond:2008aq} and form factors of $\cT_2$ \cite{vanNeerven:1985ja, Gehrmann:2011xn, Brandhuber:2012vm}.%
\footnote{This result disagrees with the findings of \cite{Bork:2010wf}, where a different result for the collinear anomalous dimension was obtained, see Section \ref{remainder-constants} for more details. } 

Finally, we use the symbol of transcendental functions and the related, refined notion of the coproduct  \cite{Golden:2014xqa}
to construct the remainders in an extremely compact form.
For the remainder of $\cF_{\cT_3,3}$ we find a three-line expression containing only classical
polylogarithms, while the answer for $\cF_{\cT_k,k}$ is a combination of universal, compact building blocks which contain classical polylogarithms supplemented by just two Goncharov polylogarithms.

We end  with a brief outline of the rest of the paper. After a short summary of form factors of $\cO_k$ and $\cT_k$ in Section \ref{summary}, we construct the minimal form factor of $\cT_3$ at two loops
in terms of a basis of integral functions in Section \ref{sec:T3}. 
In Section \ref{sec:remainder} we define finite remainder functions of the minimal two-loop form
factors, and use the powerful concept of the symbol of transcendental functions to rewrite the result in terms of classical polylogarithms only. 
In Section \ref{sec:allk} we work out the analytic results for 
form factors of $\cT_k$ with $k>3$ and are able to express them in terms of a single, universal building block that depends on three scale-invariant ratios of Mandelstam variables. Again, using the symbol and coproduct of transcendental functions we find a compact answer which, in addition to classical polylogarithms,  contains also two Goncharov polylogarithms. 
Finally, in Section \ref{sec:discussion} we analyse in some detail the behaviour of form factors in collinear and soft limits, and note that minimal form factors have unconventional factorisation properties compared to amplitudes and non-minimal form factors.

\section{Summary of tree-level and one-loop  form factors}
\label{summary}

We begin this section with a brief review of supersymmetric form factors at tree level and one loop. This will allow us to fix conventions and to introduce the building blocks which enter the unitarity cuts at two loops performed  in Sections  \ref{sec:T3} and \ref{sec:allk}. The reader more interested in the results can skip this section and move directly to the next one. 

\subsection{Super form factors of half-BPS operators}

A form factor of a local, gauge-invariant operator $\O(x)$ is defined as the Fourier transform of its matrix element taken between the vacuum and an on-shell $n$-particle state,
\begin{equation}
\label{eq:ff-def}
F_{\O,n}(1, \ldots, n;q) \ := \   \int\!d^4x \, e^{- i q x} \  \langle 1 \cdots n |\cO (x)  |0\rangle  \ = \ \delta^{(4)} \big(q - \sum_{i=1}^n p_i\big) 
  \langle 1 \cdots   n |\cO (0)  |0\rangle\ . 
\end{equation}
While  the external states are massless and on shell, the operator carries a momentum $q$ which in general is off shell. 

\noindent
In this paper we will focus on half-BPS operators of the form
\beq
\O_k\,:=\,\Tr( \phi_{AB}^k)\ ,
\eeq
for fixed $A$ and $B$, with  
$k\geq 3$ (the case $k=2$ was discussed in \cite{Brandhuber:2011tv}).

The operator $\cO_k$ is part of a half-BPS supermultiplet of operators, and is nothing but a simple generalisation of the chiral part of the stress-tensor multiplet \cite{Eden:2011yp,Eden:2011ku}. We  will denote the corresponding supersymmetric operator by $\T_k(x,\theta^+)$.  It depends on half of the fermionic coordinates of the $\N=4$ harmonic superspace  \cite{Galperin:1984av,Galperin:2001uw}, denoted collectively by~$\theta^\pm$,
\begin{align}
\label{eq:theta+-}
\theta^+\ \leftrightarrow \ \theta^{+a}_{\alpha},\ \,\qquad a&=1,2\, ,\qquad \alpha=1,2\ , 
\nonumber \\
\theta^-\ \leftrightarrow \ \theta^{-a'}_{\dot\alpha},\qquad a'&=1,2\, ,\qquad \dot\alpha=1,2\ .
\end{align}
Instead of taking the $k^{\rm th}$  power of a scalar field $\phi_{AB}$, we now take the $k^{\rm th}$ power of a particular projection $W^{++}$ of the chiral vector multiplet $W_{AB}$,%
\footnote{For details on the harmonic projections mentioned above, we refer the reader to 
\cite{Eden:2011yp} and  to Section 2 of \cite{Penante:2014sza}. }
\begin{align}
\label{eq:tk}
\T_k (x, \theta^{+}) \, :=  \, \Tr \big[( W^{++}(x,\theta^+))^k\big]  \,=\,\Tr \big[( \phi^{++}(x))^k\big] \, + \, \cdots 
\ ,
\end{align}
where the ellipses stand for other operators related to $\Tr \big[( \phi^{++})^k\big]$ via supersymmetry transformations and which are accompanied by higher powers of the fermionic variables $\theta^+$. 
In what follows, we will focus on the lowest component, or equivalently the $(\theta^+)^0$  component of $\T_k$ in \eqref{eq:tk},  which is nothing but $\O_k$. 
By choosing an appropriate harmonic projection \cite{Eden:2011yp} one can simply replace $\phi^{++}$  by
$\phi_{12}$ without loss of generality. This supersymmetric notation allows us to make supersymmetry Ward identities manifest,  and relates form factors of the lowest component operator to form factors of all the other component operators appearing in $\T_k$.

Super form factors are defined in a similar way as their bosonic counterparts \eqref{eq:ff-def}, except that now one takes a super  Fourier transform. The super form factors of $\T_k$ are given by
\beq
\label{eq:superff-def}
\cF_{\cT_k,n} (1, \ldots, n;  q, \gamma_{+} )
\ := \ \int\!\!d^4x\, d^4 \theta^{+} \
e^{-(iqx + i \theta_\alpha^{+a}
\gamma_{+a}^\alpha) } \, \langle  \, 1 \cdots n\,  | \cT_k(x, \theta^+)
\,  | 0   \ran\ \, ,
\eeq
where now the operator carries fermionic momentum $\gamma_+$ in the $\theta^+$ directions in addition to the  momentum $q$. Here the external state 
is  described using Nair's  formalism \cite{Nair:1988bq}. This is based on the introduction of a super-wavefunction 
\beq
\label{eq:supermultiplet}
\Phi(p,\eta) := g^+(p) + \eta_A \lambda^A(p) + {\eta_A \eta_B\over
2!} \phi^{AB}(p) + \epsilon^{ABCD} {\eta_A \eta_B \eta_C\over 3!}
\bar\lambda_D(p) + \eta_1 \eta_2 \eta_3 \eta_4 g^-(p)
\, ,
\eeq
where  we denote by $\big(g^+(p), \ldots , g^{-}(p)\big)$ the annihilation operators of the corresponding particles. Here  $\eta_A$ is a Grassmann variable, and in order to extract the contribution where the helicity of particle $i$ is $h_i$, one picks the coefficient of  $(\eta_i)^{2 - 2h_i}$ of the corresponding superamplitude (or super form factor).

When $k=2$, \eqref{eq:tk} is the chiral part of the stress tensor multiplet, whose form factors were studied in \cite{Brandhuber:2010ad, Brandhuber:2011tv}. For general $k$, these form factors were studied  in \cite{Penante:2014sza} at tree level and one loop. Here we will  extend this work and explore form factors of $\T_k$ at two loops. Therefore it is useful to summarise some relevant results form  \cite{Penante:2014sza} which will appear throughout our calculations.

\subsection{Tree-level results }

It is known that the MHV super form factors of the chiral part of the stress tensor multiplet $\T_2$ are given by a  simple expression \cite{Brandhuber:2011tv}, which is reminiscent of the Parke-Taylor formula for MHV superamplitudes \cite{ParkeTaylor:1986, Mangano:1987xk},
\begin{equation} 
\label{eq:stresstensorFF}
\F^{\text{MHV}}_{\T_2,n}(1,\ldots,n;q,\gamma_+) =\frac{  \delta^{(4)}\big(q-\sum\limits_{i=1}^n\lambda^i\tl^i\big) \delta^{(4)}\big(\gamma_+-\sum\limits_{i=1}^n\lambda_i\eta_{+,i}\big) \delta^{(4)}\big(\sum\limits_{i=1}^n\lambda_i\eta_{-,i}\big)}{\la 12\ra\la 23\ra\cdots\la n1\ra}\ , 
\end{equation}
where $\eta_{\pm}$ are projections of the $\eta_A$ variables along the $\theta^{\pm}$ directions.  Since the super-operator $\T_k (x, \theta^{+})$ does not depend on the $\theta^{-}$ coordinates, 
it carries no supermomentum in these directions, which is reflected in the absence of the conjugate fermionic variable $\gamma_-$ in the last delta function.

Similarly  to the case of amplitudes,  MHV  super form factors are the simplest class of form factors
and are singled out by the fact that they carry the lowest possible fermionic degree,  which -- for the MHV super form factors of  $\T_k$ -- is $8+2(k-2)$.  These form factors are  related via supersymmetry to the form factor of $\O_k$ with an external state consisting of $k$ scalars and $(n-k)$ positive-helicity gluons. For $\T_3$, all $n$-particle MHV super form factors at tree level are given by a slight modification of \eqref{eq:stresstensorFF} \cite{Penante:2014sza},
\begin{align}
\label{eq:FF3MHVsusy-simple}
\begin{split}
\F^{\text{MHV}}_{\cT_3,n}(1,\dots,n;q,\gamma_+) \ =\ \F^{\text{MHV}}_{\T_2,n}(1,\ldots,n;q,\gamma_+)
\, \Big(\sum\limits_{i\leq j=1}^{n-2}(2-\delta_{ij})\dfrac{\b{n\,i}\b{j\,n-1}}{\b{n-1\,n}}\eta_{-, i}\cdot\eta_{-, j}\Big)\, ,
\end{split}
\end{align}
where we have introduced the shorthand notation  $\eta_{-, i}\cdot\eta_{-,j}\, :=\, \frac{1}{2}\, \eta_{-a, i}\eta_{-b, j}\, \epsilon^{ab}$. Note that the extra factor increases the fermionic degree of \eqref{eq:stresstensorFF} by two units, which in turn is necessary to encode the larger $R$-charge carried by the operator. Tree-level MHV form factors for arbitrary  $k>3$ were  also derived in
\cite{Penante:2014sza} using a slightly modified version of the BCFW recursion relations. 
The  result is a generalisation of \eqref{eq:FF3MHVsusy-simple}, 
and the interested reader can find it  in Eqn.~$(4.12)$ of \cite{Penante:2014sza}. 

\subsection{One-loop results }

The one-loop MHV form factors of the  operators $\T_k$ for general $k$ have been   derived in \cite{Penante:2014sza}, where it was found that they can be expressed as sums of infrared divergent and finite contributions. Taking  $k=3$ as an example,%
\footnote{MHV form factors of general operators $\T_k$ can be found in Eqn.~$(6.19)$ of \cite{Penante:2014sza}.} 
we have
\begin{align}
\begin{split}
\label{eq:1-loop-MHV-BIS}
\F_{\T_3,n}^{\text{MHV} (1)} \ &=  \  \F_{\T_3,n}^{\text{MHV}(0)}\,    \sum_{i=1}^n {  ( - s_{i\,i+1})^{- \epsilon}   \over \epsilon^2}  \\
   &+ \F_{\T_2,n}^{\text{MHV}(0)} \sum_{a, b} f_{\T_3}(a+1,\dots,b-1,b,a)  \text{Fin}^{\rm 2me}(p_a, p_b, P, Q)\, ,
\end{split}
\end{align}
where $s_{ij}:= (p_i+p_j)^2$. Here $\text{Fin}^{\rm 2me}(p_a, p_b, P, Q)$ is the finite part of the two-mass easy box function with massless corners with momenta $p_a$ and $p_b$,  and massive corners with momenta $P$ and $Q$. The coefficients $f_{\T_3}$ are given by
\begin{align}
f_{\T_3}(a+1,\dots,b-1,b,a) \,=\, \sum\limits_{i\leq j=a+1}^{b-1}(2-\delta_{ij})\dfrac{\b{a\,i}\b{j\,b}}{\b{b\,a}}\eta_{-, i}\cdot\eta_{-, j}\, .
\end{align}
The result 
\eqref{eq:1-loop-MHV-BIS} can also be graphically represented as
\begin{equation} \label{Fig:tribox}
\includegraphics[width=0.7\textwidth]{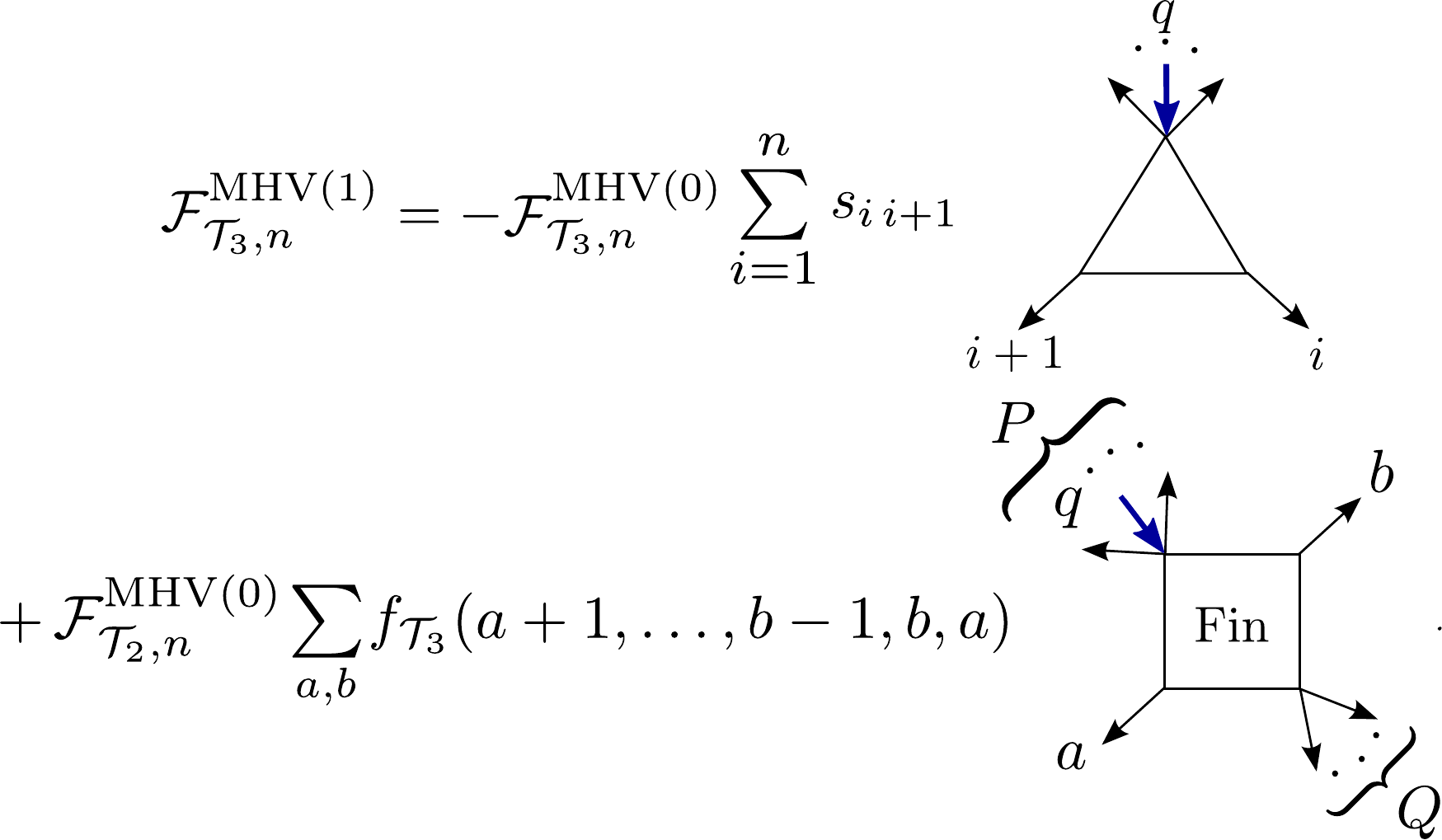}
\end{equation}
Our two-loop calculation will focus on minimal form factors, $\F_{\T_k,k}$,  where the number of external legs is identical to that of fields in the operators.  
At one loop, the minimal form factors are given simply by a sum of  infrared-divergent terms:
\beq
\label{eq:1-loop-MHV-all-k}
\F_{\T_k,k}^{(1)}\ =   \    -\F_{\T_k,k}^{(0)} \sum_{i=1}^k s_{i\,i+1} I_{3}^{1{\rm m}}(p_i,p_{i+1}; p_i+p_{i+1}) \ = \
\F_{\T_k,k}^{(0)}\, 
\sum_{i=1}^k 
{  ( - s_{i\,i+1})^{- \epsilon}   \over \epsilon^2} \, ,   
\eeq
where $I_{3}^{1{\rm m}}$ is the one-mass triangle integral indicated in the first line of \eqref{Fig:tribox}.

\noindent In the following we will be dealing only with form factors of the half-BPS operators $\T_k$ and $\O_k$, and we find  it convenient to introduce a more concise notation,
\begin{align*}
\F_{k,n}\ &\leftrightarrow\ \F_{\T_k,n}\qquad\, \text{Super form factor of } \T_k \ , \\
F_{k,n}\ &\leftrightarrow\ F_{\O_k,n}\qquad\text{Form factor of } \Tr\big[(\phi_{12})^k\big]\ .
\end{align*}
Notice that in order to have  a non-vanishing $F_{k,k}$, all external states must be equal to $\phi_{12}$.
 
\subsection{Colour decomposition and planarity}
\label{sec:colour-planar}

In this section we briefly consider the colour decomposition of form factors and its implications for the calculation of  two-loop  form factors of $\T_k$. 

Following the same procedure  as for scattering amplitudes, a planar $n$-point  form factor of a certain  single-trace operator $\O$ can be expressed as
\begin{equation} \label{eq:colordecomp}
\F^{\, a_1\cdots a_n}_{\O,n}=\sum_{\sigma\in S_n/\mathbb{Z}_n}\mathrm{Tr}(T^{a_{\sigma(1)}}T^{a_{\sigma(2)}} \cdots T^{a_{\sigma(n)}})\F_{\O,n}(\sigma(1),\sigma(2),\dots, \sigma(n);q,\gamma_+)\ ,
\end{equation}
where $T^a$ are  fundamental generators of  $SU(N)$, and the  $\F_{\O,n}$  are colour-ordered form factors. 

An important remark is in order here. For the case of $\T_2$, the minimal ({\it i.e.}~two-point) form factor has the colour factor $\Tr(T^a T^b)$, which is simply $\delta^{ab}$. As noticed in \cite{Brandhuber:2012vm}, this simple fact has striking consequences for the two-loop calculations. Consider for instance a two-particle cut of the form 
\beq
\int\!d{\rm LIPS} (\ell_1, \ell_2; p_1 + p_2 ) \ \F_{2,2}^{(0)}(\ell_1,\ell_2;q,\gamma_+) \times \A^{(1)}_4(-\ell_1,-\ell_2, 1,  2)\ .
\eeq
The colour factor $\delta^{a_{\ell_1} a_{\ell_2}}$ arising from  the form factor can  contract with a double-trace term from the complete  one-loop amplitude $\A^{(1)}_4$,  generating extra powers of $N$. Hence, these double-trace terms, which are normally subleading in colour, are lifted to leading order in $N$. As a consequence, one has to keep double-trace contributions from  $\A^{(1)}_4$. 
This is the reason why planar two-loop form factors of $\T_2$ receive contributions from non-planar integral topologies \cite{vanNeerven:1985ja}. In \cite{Brandhuber:2012vm} it was shown that this also applies to non-minimal form factors of $\T_2$.

Remarkably, this is not the case for $k>2$ at two loops. This  is because now one can only have three- or higher-point form factors entering the cuts, which are never dressed with $\delta^{ab}$ colour factors. This  situation is very similar to the case of planar scattering amplitudes, where only planar integrals contribute.
We will make use of this fact in the two-loop calculation of $\F_{k,k}$ in the following sections.

Note that form factors are still intrinsically non-planar quantities, because the operators are colour singlets. In particular, we have checked that non-planar integrals will  appear at three loops in the expression of  the form factors of $\T_k$ with $k>2$. Finally, we note that even for one- and two-loop form factors,  where only planar integrals contribute, one cannot define a unique set of region momenta for all integrals contributing to a certain form factor.

\section{Minimal form factor of $\T_3$ at two loops}
\label{sec:T3}
\subsection{Unitarity cuts}
In this section we calculate the two-loop form factor $F_{3,3}^{(2)}$ using
generalised unitarity. 
In particular we show that two-particle cuts combined with two different types of three-particle cuts are sufficient to fix the result uniquely and express it as a linear combination of planar two-loop master integrals.

\noindent We start by considering the two-particle cuts. 
\begin{figure}[htb]
\centering
\includegraphics[width=0.8\textwidth]{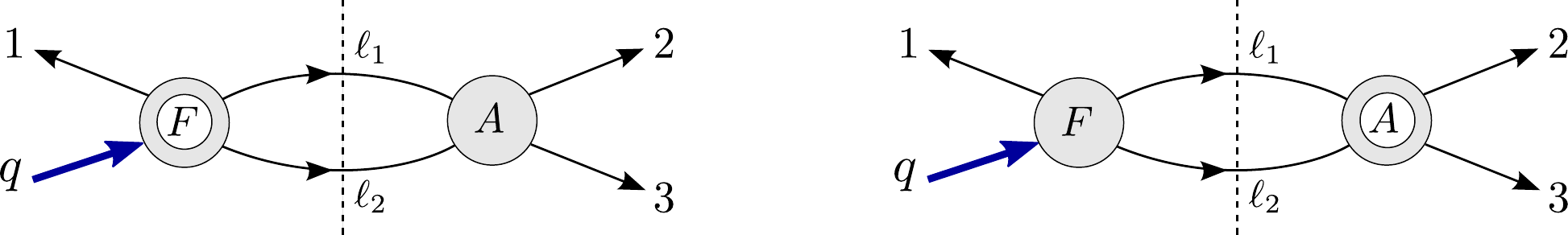}
\caption{\it Two-particle cuts of $F^{(2)}_{3,3}$ in the kinematic channel $s_{23}=(p_2+p_3)^2$. There are two possible factorisations: $F^{(1)} \times A^{(0)}$ (left) or $F^{(0)} \times A^{(1)}$ (right). Cyclic permutations of the legs 123 generate the remaining two-particle cuts.}
\label{fig:twoloopcut1}
\end{figure}
\noindent

At two-loop level there are two such cuts as shown in Figure \ref{fig:twoloopcut1}.
First, we consider the cut on the left-hand side of  Figure \ref{fig:twoloopcut1}, where a one-loop form factor is merged with a tree-level four-point amplitude. The cut integrand is given by
\beq
\mathcal{C}_{s_{23}}^{(1)} \, = \, \int  \! d\text{LIPS}(\ell_1,\ell_2;P)\, F^{(1)}_{3,3}(1, \ell_1, \ell_2; q) \, A^{(0)}_4(-\ell_2, -\ell_1, 2, 3)
\ , 
\eeq
where $P= p_2+p_3$ and, making the helicities explicit,
\beqa \label{cut1}
A^{(0)}_4(-\ell_2^{\phi_{34}}, -\ell_1^{\phi_{34}}, 2^{\phi_{12}}, 3^{\phi_{12}}) &=&
{\b{ \ell_2 \, \ell_1 } \b{ 23 }  \over \b{3 \, \ell_2 } \b{ \ell_1 \, 2 } }
= {s_{23} \over 2 (\ell_1 \cdot p_2) } \, ,\\[5pt] \label{cut12}
F^{(1)}_{3,3}(1^{\phi_{12}}, \ell_1^{\phi_{12}}, \ell_2^{\phi_{12}}; q) &=&
s_{23} \, I_{3}^{1 \rm m}(\ell_1, \ell_2; p_2+p_3) + (q- \ell_2)^2\, I_{3}^{1\rm m}(p_1, \ell_1; q- \ell_2) \nonumber 
\\ [5pt]
&+& (q- \ell_1)^2\, I_{3}^{1 \rm m}(\ell_2, p_1; q-\ell_1) \, .
\eeqa
Here $I_{3}^{1 \rm m}(a, b; c)$ is a  one-mass triangle integral, 
\begin{equation*}
\includegraphics[width=0.35\linewidth]{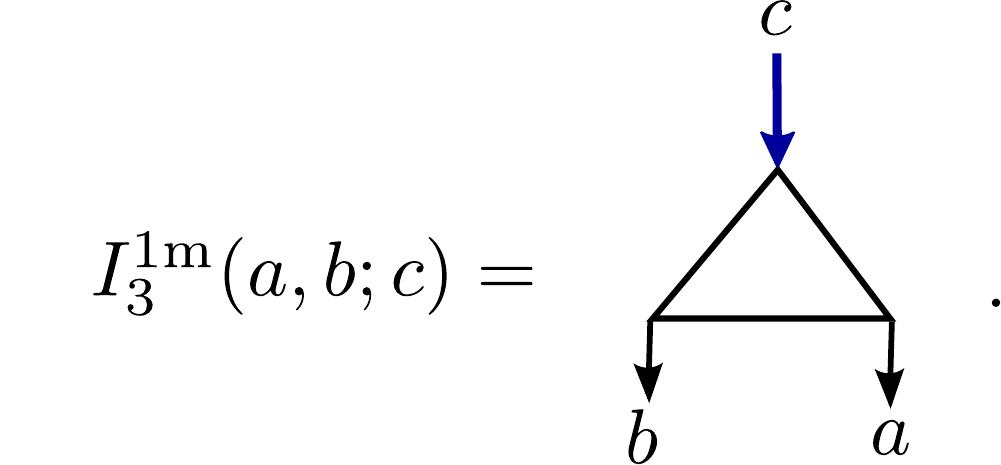}
\end{equation*}
From \eqref{cut1}, it is clear that the effect of $A^{(0)}_4(-\ell_2^{\phi_{34}}, -\ell_1^{\phi_{34}}, 2^{\phi_{12}}, 3^{\phi_{12}})$ is simply to attach the following three-propagator object with numerator $s_{23}$ to the one-loop form factor: 
\beq
\label{eq:attach}
\vcenter{\hbox{\includegraphics[width=0.16\linewidth]{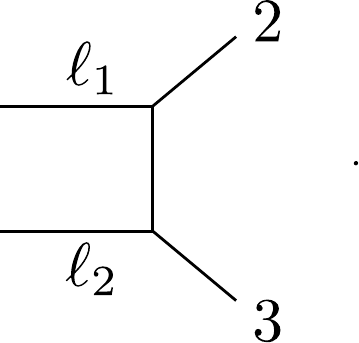}}} 
\eeq
By attaching the structure in \eqref{eq:attach}  to all the triangles appearing in the one-loop form factor \eqref{cut12}, we find that the cut integrand $\cC_{s_{23}}^{(1)}$ is given by the following sum, 
\beq
\label{eq:integrals-cut-1}
\vcenter{\hbox{\includegraphics[width=0.95\linewidth]{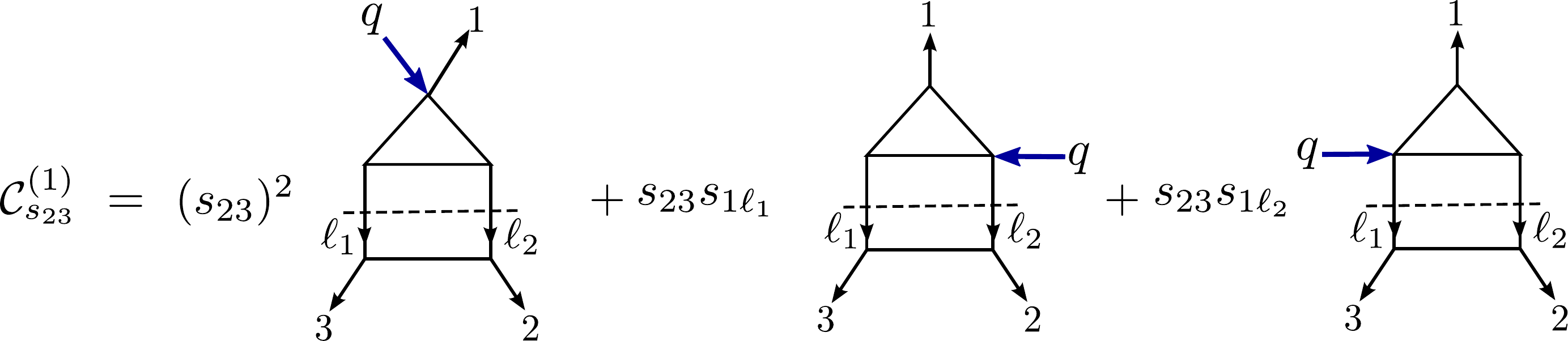}}}\, .
\eeq\\[3pt]
The straight dashed lines in the integrals above indicate that the momenta $\ell_1$ and $\ell_2$ are cut. We now introduce a more concise notation for numerators which will be used in the following. To indicate a factor of $s_{ij} $ in the numerator of an integral, we draw a curved dashed line passing through two propagators whose momenta sum to $p_i+p_j$. In this notation, \eqref{eq:integrals-cut-1} can be represented as 
\beq
\label{eq:integrals-cut-2}
\vcenter{\hbox{\includegraphics[width=0.6\linewidth]{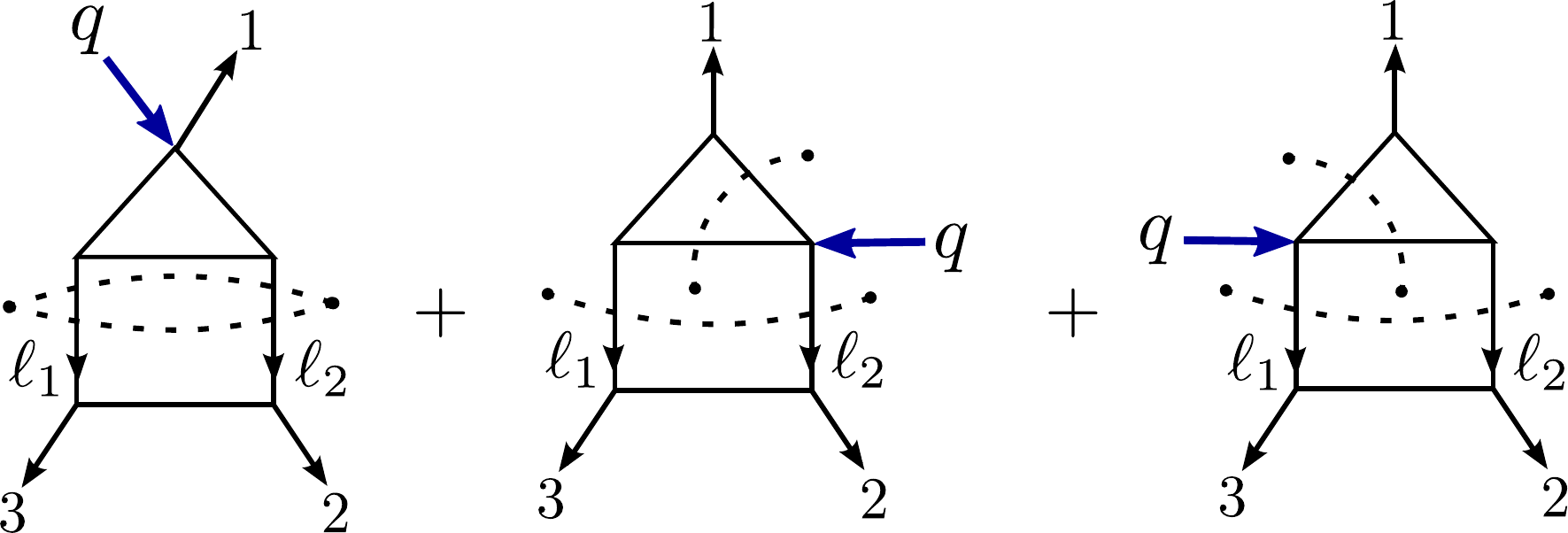}}}\,.
\eeq
Note that at this stage we have also uplifted the cut integrals to full Feynman integrals by replacing the cut legs by propagators.
We stress that this procedure induces ambiguities in the numerators, since on the cut $\ell_1^2=\ell_2^2=0$, and hence we cannot distinguish $s_{1\,\ell_1}$ from  $2\, p_1 \cdot \ell_1$ or $s_{1\,\ell_2}$ from $2\, p_1 \cdot \ell_2$. Such ambiguities will be eliminated later using
triple cuts.

\noindent The second two-particle cut, depicted on the right-hand side of Figure \ref{fig:twoloopcut1}, is given by
\beq
\mathcal{C}_{s_{23}}^{(2)} \, = \, \int \! d\text{LIPS}(\ell_1,\ell_2;P)\, F^{(0)}_{3,3}(1, \ell_1, \ell_2; q) \, A^{(1)}_4(-\ell_2, -\ell_1, 2, 3)
\ , 
\eeq
 where
\beqa
F^{(0)}_{3,3}(1^{\phi_{12}}, \ell_1^{\phi_{12}}, \ell_2^{\phi_{12}}; q) &=&
1\, , \\[5pt]
A^{(1)}_4(-\ell_2^{\phi_{34}}, -\ell_1^{\phi_{34}}, 2^{\phi_{12}}, 3^{\phi_{12}}) &=&
{s_{23} \over 2\,(\ell_1 \cdot p_2) } \big[ s_{23}\, (p_2-\ell_1)^2\, I_{4}^{0\rm m} (-\ell_2, -\ell_1, 2, 3) \big] \, ,
\eeqa
where $I_{4}^{0\rm m}$ stands for the zero-mass scalar box integral, 
\begin{equation}
\includegraphics[width=0.35\textwidth]{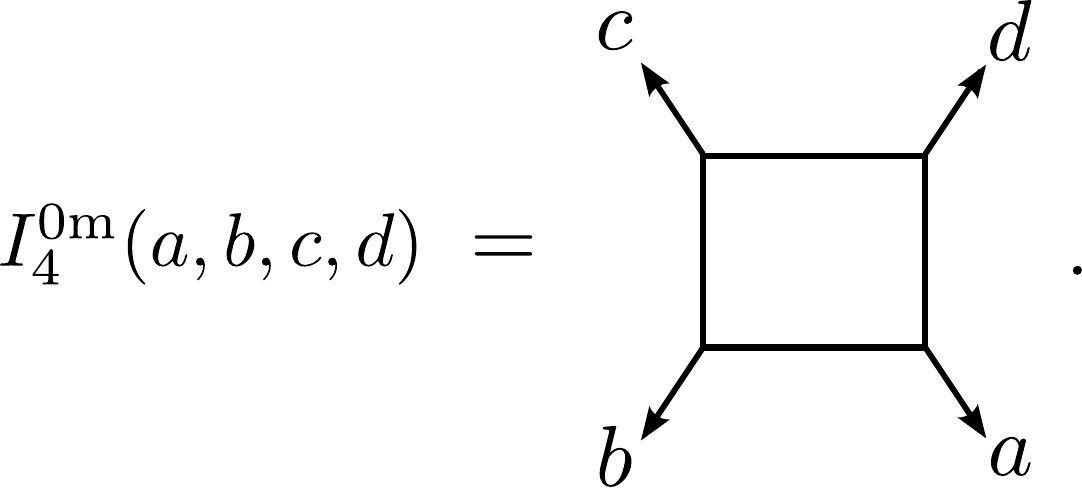}
\end{equation}
Uplifting $\mathcal{C}_{s_{23}}^{(2)}$ to a full Feynman integral we obtain the contribution depicted in \eqref{eq:integrals-cut-2-2}, 
\beq
\label{eq:integrals-cut-2-2}
\vcenter{\hbox{\includegraphics[width=0.14\linewidth]{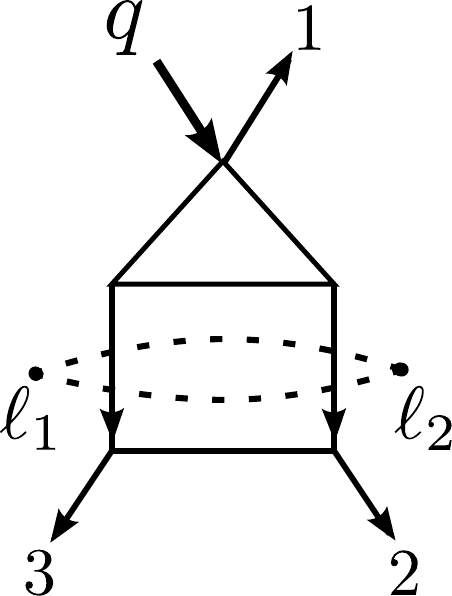}}}\ ,
\eeq
which was already detected in the first two-particle cut.
Therefore \eqref{eq:integrals-cut-2} alone comprises the full result for this cut.

\begin{figure}[h]
\centering
\includegraphics[width=0.38\linewidth]{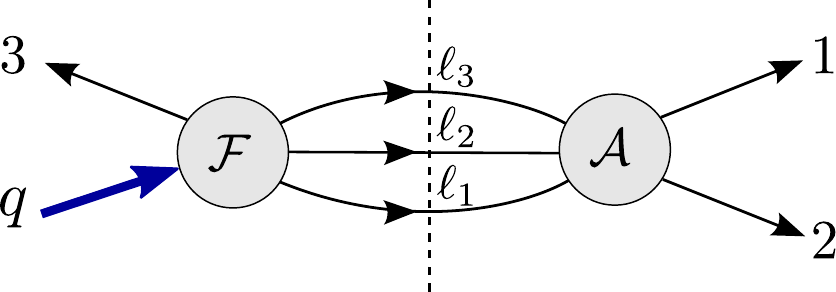}
\caption{\it A possible three-particle cut of $\F_{3,3}^{(2)}$.}
\label{fig:tricut1}
\end{figure}
We now move on to investigate three-particle cuts. The first case we want to consider is shown in Figure \ref{fig:tricut1}.
This three-particle cut is given by
\begin{equation}
\label{eq:triplecut1}
\cC^{(3)}_{s_{12}}\, =\, \int \! d\text{LIPS}(\ell_1,\ell_2,\ell_3 ;P) \int d^{12}\eta \ \F^{(0)}_{3,4}(3,\ell_3,\ell_2,\ell_1;q,\gamma^+)\, \A^{(0)}_{5} (1,2,-\ell_1,-\ell_2,-\ell_3)\ ,
\end{equation}
where $P=p_1+p_2$ and $d^{12}\eta = d^4\eta_{\ell_1} d^4\eta_{\ell_2} d^4\eta_{\ell_3}$. Importantly,
in order to perform the sum over internal helicities efficiently
we use the supersymmetric formalism for form factors developed in \cite{Brandhuber:2011tv}, and adapted in \cite{Penante:2014sza} to the case of the operators $\cT_k$. At the end of the calculation we will select all external particles to be $\phi_{12}$. 

There are two distinct choices for the form factor and  amplitude participating in the cut, namely 
\begin{equation}
\F^{\rm NMHV}_{3,4}\times \A^{\rm MHV}_5\quad \text{and} \quad \F^{\rm MHV}_{3,4} \times \A^{\overline{\rm MHV}}_5 \ .
\end{equation}
We consider first the case $\F^{\rm NMHV} \times \A^{\rm MHV}$. The tree-level expressions entering \eqref{eq:triplecut1} are given by (omitting  a trivial delta function of momentum conservation)\\
\begin{align}
\label{eq:FFNMHV}
\begin{split}
\F^{\text{NMHV}}_{3,4}
&=\delta^{(4)}\big(\gamma_+ - \sum_i \lambda_i \eta_{+, i} \big) \delta^{(4)}
\big(\sum_i \lambda_i \eta_{-, i} \big)
\Bigg[
{\delta^{(4)}\big( [\ell_3\,\ell_2]\eta_3 + [\ell_2\,3] \eta_{\ell_3} + [3\,\ell_3] \eta_{\ell_2} \big)
(\eta_{-,1})^2 \over [\ell_3\,\ell_2][\ell_2\,3][3\,\ell_3] (s_{3\ell_3\ell_2})^2 } 
 \\ 
&- (\ell_1 \leftrightarrow \ell_3 )
\Bigg] \, ,
\end{split} \\
\A^{\rm MHV}_5 & =\frac{ \delta^{(8)}\big( \lambda_1 \eta_1 + \lambda_2 \eta_2 -\lambda_{\ell_1} \eta_{\ell_1} -\lambda_{\ell_2} \eta_{\ell_2}-\lambda_{\ell_3} \eta_{\ell_3}\big)}{\b{12}\b{2\,\ell_1}\b{\ell_1\,\ell_2}\b{\ell_2\,\ell_3}\b{\ell_3\,1} }\ ,
\end{align}\\
where the NMHV form factor of $\cT_3$ is given in Eqn.~(4.8) from \cite{Penante:2014sza}.
After performing the integrations over the internal $\eta_{\ell_i}$'s, we arrive at the result 
\begin{align}
\label{eq:res-triplecut-1}
\left. 
\cC_{s_{12}}^{(3)}\right|_{\rm A}\,=\,\F_{3,3}^{(0)} \frac{\b{12}}{\b{2\,\ell_1}\b{\ell_1\,\ell_2}\b{\ell_2\,\ell_3}\b{\ell_3\,1}} \left(\frac{[3|q|\ell_1\ra^2}{[\ell_3\,\ell_2][\ell_2\,3][3\,\ell_3]}-  (\ell_1 \leftrightarrow \ell_3 )\right)\, .
\end{align}
The second case is $\F^{\rm MHV} \times \A^{\overline{\rm MHV}}$. The expressions entering \eqref{eq:triplecut1} can be  written as\\
\begin{align}
\F^{\rm MHV}_{3,4} &= \frac{1}{\b{3\,\ell_3} \b{\ell_3\,\ell_2}\b{\ell_2\,\ell_1}\b{\ell_1\, 3} } \left[(\eta_{3,-})^2\frac{\b{3\,\ell_3}\b{\ell_1\,3}}{\b{\ell_3\,\ell_1}}- (\eta_{\ell_2,-})^2\frac{\b{\ell_2\,\ell_3}\b{\ell_1\,\ell_2}}{\b{\ell_3\,\ell_1}}\right]\ ,\\[5pt]
\A^{\overline{\rm MHV}}_5 &= \frac{1}{[2\,\ell_1][\ell_1\,\ell_2][\ell_2\,\ell_3][\ell_3\,1][12]^9} \prod_{i=1}^3 \, \delta^{(4)}\big([12]\,\eta_{\ell_i}+  [2\,\ell_i]\, \eta_1 +  [\ell_i\,1] \,\eta_2 \big)\ .
\end{align}\\
After summing over internal helicities, we get
\begin{align}
\label{eq:res-triplecut-2}
\begin{split}
\left. 
\cC_{s_{12}}^{(3)}\right|_{\rm B}  \,=\,\F_{3,3}^{(0)}\, &  \frac{(s_{12})^2 [12]}{\b{3\,\ell_3} \b{\ell_3\,\ell_2} \b{\ell_2\,\ell_1}\b{\ell_1\, 3}[2\,\ell_1][\ell_1\,\ell_2][\ell_2\,\ell_3][\ell_3\,1] } \\[8pt]
\times\, & \left(\frac{\b{3\,\ell_3}\b{\ell_1\,3}}{\b{\ell_3\,\ell_1}}-\frac{[\ell_2|q|3\ra ^2 }{(s_{12})^2}\frac{\b{\ell_2\,\ell_3}\b{\ell_1\,\ell_2}}{\b{\ell_3\,\ell_1}} \right)\ .
\end{split}
\end{align}\\
Summarising, the total result for the cut \eqref{eq:triplecut1} is the sum of \eqref{eq:res-triplecut-1} and \eqref{eq:res-triplecut-2},
\begin{align}
\label{3pcab}
\cC_{s_{12}}^{(3)}\,=\,\left. \cC_{s_{12}}^{(3)}\right|_{\rm A} + \left. \cC_{s_{12}}^{(3)}\right|_{\rm B}\ .
\end{align}
Taking the purely scalar component of this cut amounts simply to performing the replacement $\F_{3,3}^{(0)} \to 1$. 
\begin{figure}[h]
\centering
\includegraphics[width=0.42\linewidth]{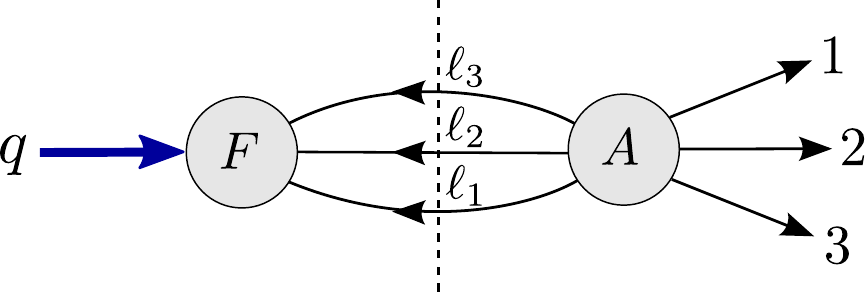}
\caption{\it A second possible three-particle cut of $\F_{3,3}^{(2)}$.}
\label{fig:tricut2}
\end{figure}

The next cut we wish to consider is shown in Figure \ref{fig:tricut2},  and is given by
\begin{equation}
\label{eq:triplecut2}
\cC_{s_{123}}^{(4)} \, =\, \int \! d\text{LIPS}(\ell_1,\ell_2,\ell_3 ;-q)\, \int\!d^{12} \eta \, \cF^{(0)}_{3,3}(-\ell_1,-\ell_2,-\ell_3;q)\, \cA^{{\rm NMHV }(0)}_{6} (1,2,3,\ell_1,\ell_2,\ell_3)\ .
\end{equation}
Selecting the external particles to be all scalars $\phi_{12}$, we see that the only non-vanishing form factor contributing to the cut is  $F^{(0)}_{3,3} (-\ell_1^{\phi_{12}},-\ell_2^{\phi_{12}},-\ell_3^{\phi_{12}};q) = 1$ (again omitting a momentum conservation delta function). 
This is the only internal helicity assignment we need to consider, thus the single amplitude appearing on the right-hand side of the cut is the following six-scalar NMHV amplitude,  
\begin{align}
\begin{split}
A^{\rm NMHV}_6(1^{\phi_{12}},2^{\phi_{12}},3^{\phi_{12}},\ell_1^{\phi_{34}},\ell_2^{\phi_{34}},\ell_3^{\phi_{34}})\, = \, &\frac{\b{\ell_2\,\ell_3}[23]\la 1 | \ell_2+\ell_3|\ell_1]}{\b{\ell_3\,1}[3\,\ell_1]\la \ell_2|\ell_3+1|2]s_{1\ell_2\ell_3}}\\[5pt]
+\, & \frac{\b{\ell_1\,\ell_2}[12]\la 3 | \ell_1+\ell_2|\ell_3]}{\b{3\,\ell_1}[\ell_3\,1] \la \ell_2|\ell_1+3|2]s_{3\ell_1\ell_2}}\ . 
\end{split}
\end{align}
Hence the result of this triple cut is simply given by   
\beq
\label{tcc}
\cC_{s_{123}}^{(4)} \, =\, \int \! d\text{LIPS}(\ell_1,\ell_2,\ell_3 ;-q)\,A^{\rm NMHV}_6(1^{\phi_{12}},2^{\phi_{12}},3^{\phi_{12}},\ell_1^{\phi_{34}},\ell_2^{\phi_{34}},\ell_3^{\phi_{34}})
\ . 
\eeq

\subsection{Two-loop result}
\label{sec:result}
The two-particle cuts employed earlier show that the full two-loop result contains the combination of integrals  \eqref{eq:integrals-cut-2}. As discussed earlier, this set of cuts does not uniquely determine the numerators of these integrals, and furthermore does not probe the presence of any integral function which  only has  three-particle cuts. 

Using the result of the three-particle cuts \eqref{3pcab} and \eqref{tcc}, we can fix all such ambiguities. In particular, we have identified two additional integral topologies without two-particle cuts contributing to the final result. 
The unique function with the correct two- and three-particle cuts turns out to be
\beq \label{eq:basisT3}
\cF^{(2)}_{3,3} = \sum^3_{i=1} \Big[  I_1(i) + I_2(i) + I_3(i) +I_4(i) - I_5(i)  \Big]  \, ,
\eeq
where the integrals $I_k$ are  given by
\beq
\label{eq:basis}
\vcenter{\hbox{\includegraphics[scale=0.50]{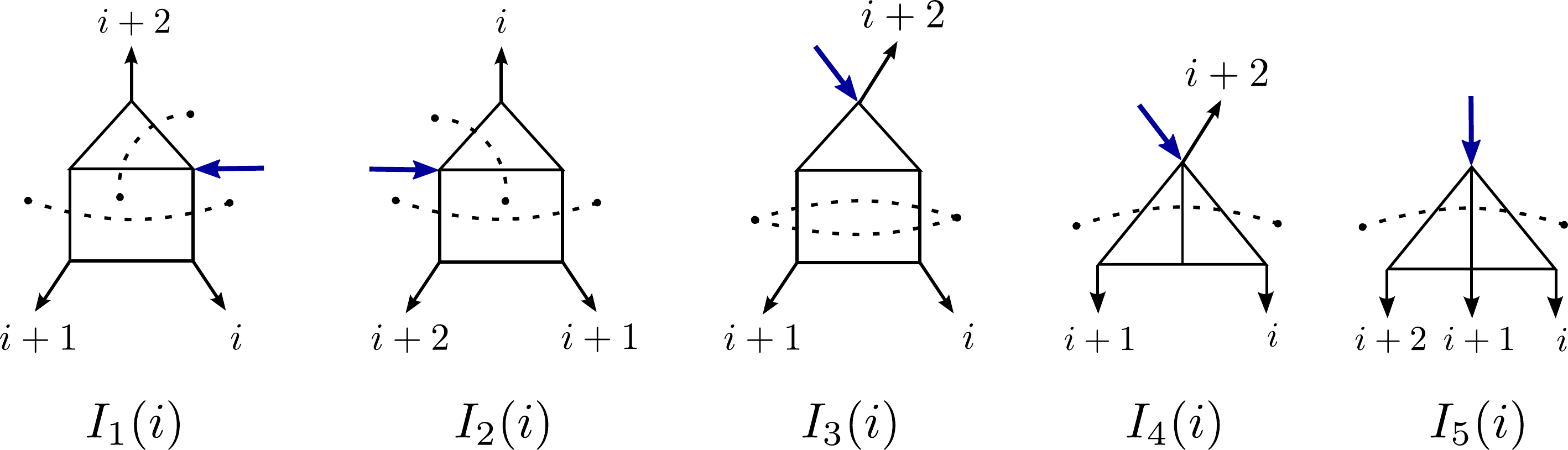}}}\, .
\eeq
Explicit expressions for most integrals that appear in \eqref{eq:basis} can be found in  \cite{Gehrmann:2000zt}. The ones that cannot be found there are $I_1$ and $I_2$, which have the same topology. As an example
we focus on $I_2$, {\it i.e.}~the second integral in \eqref{eq:basis}, 
and employ the \texttt{FIRE} algorithm \cite{Smirnov:2008iw} in order to  decompose it in terms of scalar two-loop master integrals, with the  result
\begin{equation} 
\label{uuu} 
\includegraphics[width=\linewidth]{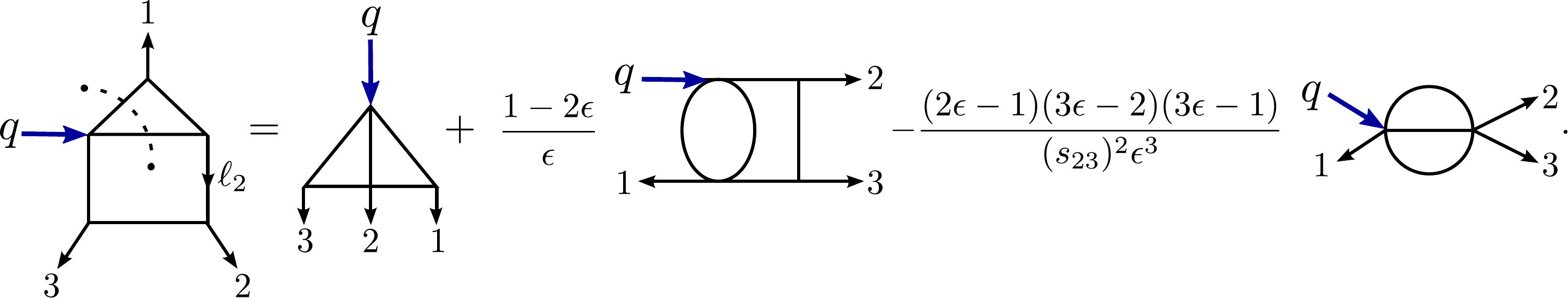}
\end{equation}
The dashed line in the integral on the  left-hand side of \eqref{uuu}  represents the numerator $s_{1 \ell_2}$ (for simplicity, we divided the whole expression by $s_{23}$ when compared to $I_2(1)$).

A few comments are in order here.

{\bf 1.} The first integral on the right-hand side of  \eqref{uuu} can naturally be combined with $I_5(1)$ in \eqref{eq:basisT3}. This is important as it ensures that the contribution to the final answer from this topology is a linear combination of multiple polylogarithms with purely numerical, {\it i.e.}~momentum-independent coefficients. The explicit expressions of the first and second integrals in terms of two-dimensional Goncharov polylogarithms can be found in \cite{Gehrmann:2000zt}, 
Eqns.~(4.32)--(4.37) and Eqns.~(4.26)--(4.31), respectively. Also note that the $\epsilon$-dependent prefactor of the second integral ensures that the expanded result has homogenous degree of transcendentality. Finally, the third integral in \eqref{uuu} multiplied with its $\epsilon$-dependent coefficient turns out be $-(1/2)\,  I_4(2)$ which follows from Eqn.~(5.15) of 
\cite{Gehrmann:1999as} which also has homogenous degree of transcendentality.

{\bf 2.} Once the reduction \eqref{uuu} is substituted into \eqref{eq:basisT3}
the final result is expressed as a linear combination of transcendental functions with numerical coefficients.
We refrain from writing explicitly the result at this stage because of its considerable length. Instead in the next section we will identify its universal infrared divergences and construct the  finite remainder function. This remainder is a transcendental function of degree four and, as we will show, can be brought to an extremely compact form that involves only classical polylogarithms.

{\bf 3.} As noted in \cite{Bork:2010wf}, the elements of the integral basis \eqref{eq:basis}  can be obtained from dual conformal integrals upon taking certain external region momenta to infinity. Consider for instance the simpler one-loop form factor, which may be obtained by taking one of the region momenta $x_i$ of a box integral to infinity,  as shown in \eqref{dualconf1},
\beq\label{dualconf1}
\vcenter{\hbox{\includegraphics[width=0.5\linewidth]{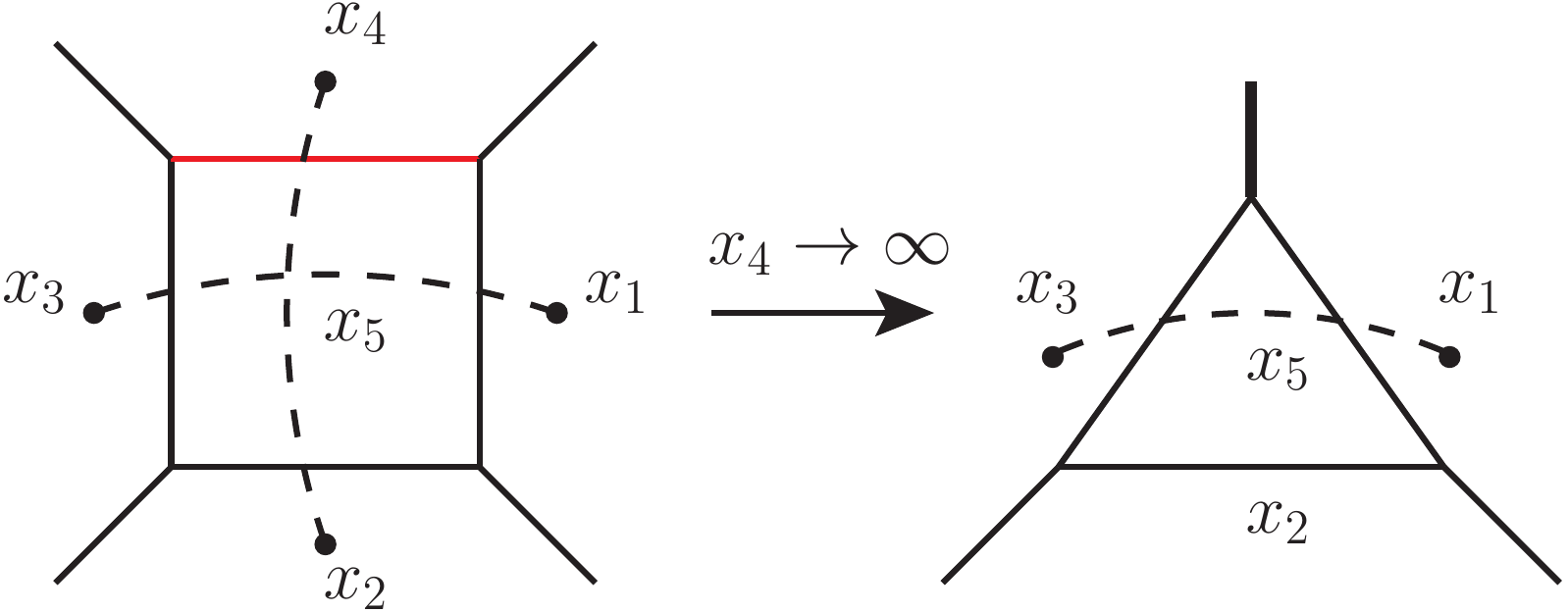}}}\, .
\eeq
In this example, as $x_4  \rightarrow \infty$, the propagator marked in red is cancelled by the numerator of the integral, and the box reduces to a triangle. If the external legs were all massive, the above two integrals would be identical due to  dual conformal symmetry \cite{Broadhurst:1993ib}. However, when there are  massless legs as in the present case, both integrals are infrared  divergent, and the symmetry is broken. Even though the symmetry is generally broken, interestingly, one could still use this ``pseudo" dual conformal symmetry to fix unambiguously 
the  numerators in each of the elements of the integral basis \eqref{eq:basis}. In what follows we show how this basis emerges from ``pseudo" dual conformal double-box and penta-box integrals,
\beq\label{dualconf2}
\vcenter{\hbox{\includegraphics[scale=0.75]{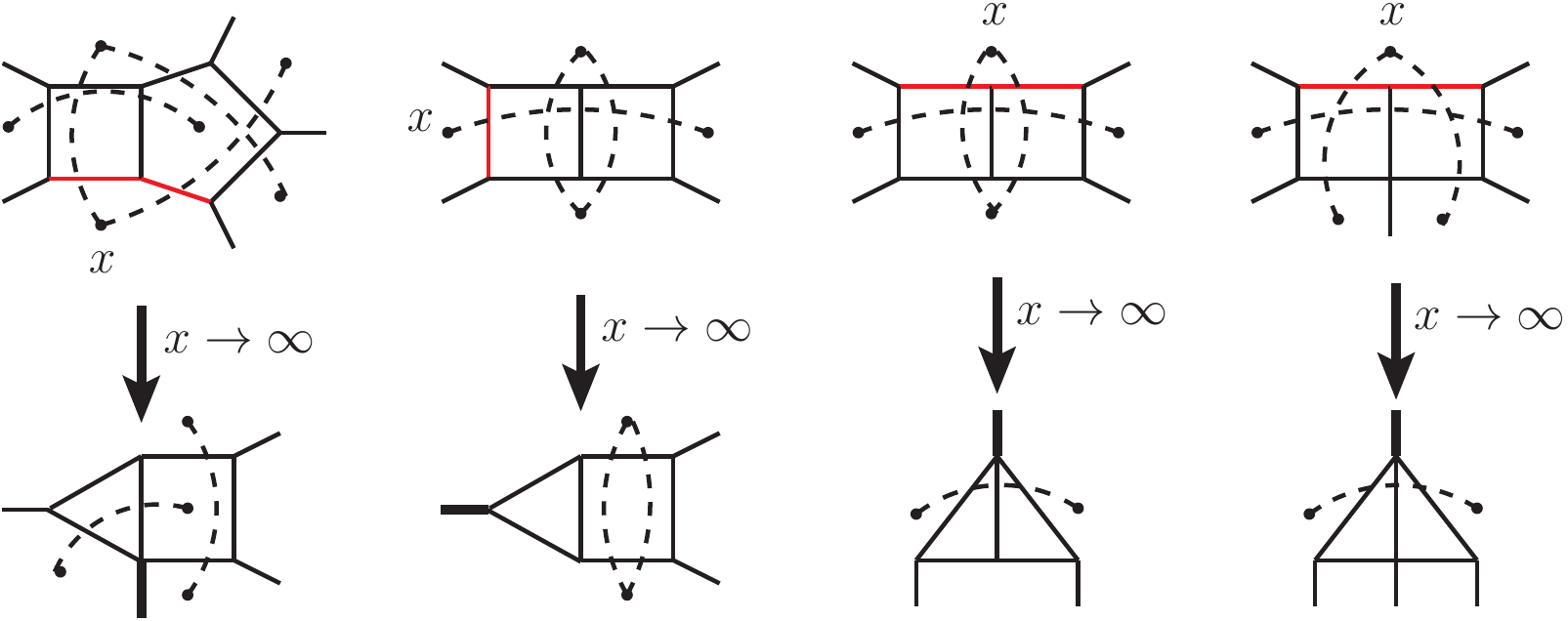}}}\, .
\eeq 
However, beyond two loops we find that not  all integrals can be obtained by using this procedure. Furthermore,  non-planar   integrals (which do not transform covariantly under the symmetry even at the integrand level) start to appear at three loops.

{\bf 4.} We observe a disagreement between our result \eqref{eq:basisT3} and the result for the same quantity as computed in Eqn.~(4.44) of \cite{Bork:2010wf}. Specifically, in our derivation the integral $G_3$ of \cite{Bork:2010wf} is missing. Our cut analysis did not detect such an integral topology and we also argue that it is  in fact not allowed  for  form factors in $\cN=4$ SYM, as it contains a triangle sub-integral which is not connected to the off-shell leg $q$, thus violating the no-triangle property of $\cN=4$ SYM, see also \cite{Boels:2012ew}.

\section{The three-point remainder function}
\label{sec:remainder}

In this section we construct a finite remainder function associated to the two-loop form factor of the operators ${\cal T}_3$, similarly to what was done in \cite{Brandhuber:2012vm} for the form factor of the stress-tensor multiplet operator ${\cal T}_2$. 
The result expressed in terms of the explicit form of the integral functions is very complicated, and in order to simplify it we determine its symbol.
From this we will finally derive a very compact form of the three-point remainder containing only classical polylogarithms. 

\subsection{Defining a form factor remainder}
\label{remainder-constants}

We begin by defining the remainder function.  Its expression is given by\footnote{In our conventions the 't Hooft coupling is defined as $a=g^2 N/(e^{\epsilon \gamma} (4 \pi )^{2-\epsilon })$. }
\beq
\label{eq:remainder}
\cR_{k,k}^{(2)} \ := \    \cG^{(2)}_{k,k}(\epsilon )\, - \, {1\over 2} \big( \cG^{(1)}_{k,k} (\epsilon) \big)^2 -  f^{(2)} (\epsilon)\  \cG^{(1)}_{k,k}  ( 2 \epsilon ) - C^{(2)}
\, + \cO (\epsilon ) \
\ , 
\eeq
where $\cG^{(L)}_{k,k}$ is the helicity-independent form factor $L$-loop ratio function, defined in the same fashion as for MHV amplitudes,
\beq
\cG^{(L)}_{k,k}\, :=\, {\F^{{\rm MHV}(L)}_{k, k}\over \F^{{\rm MHV}(0)}_{k, k}}\, ,
\eeq
and
$f^{(2)} (\eps):= f_0^{(2)} + f_1^{(2)}  \eps+ f_2^{(2)} \eps^2$.

Using \eqref{eq:basisT3} for $\F^{{\rm MHV}(2)}_{3, 3}$, we find that the $1/\epsilon^4$ and $1/\epsilon^3$ poles cancel between the first two terms of \eqref{eq:remainder}. Next
we require that the remainder is finite, and  hence that the remaining $1/\epsilon^2$ and $1/\epsilon$ poles vanish. This fixes two coefficients in the $\eps$-expansion of  $f^{(2)}$,  
\beq
f_0^{(2)} = - 2 \zeta_2 \, , \qquad f_1^{(2)} = - 2 \zeta_3 \, . 
\eeq
We note that these results for $f_0^{(2)} $ and $f_1^{(2)} $ agree with the corresponding quantities found in the case of the remainder function of the stress-tensor multiplet operator computed in 
\cite{Brandhuber:2012vm}.%
\footnote{We observe a disagreement between our result $f_1^{(2)} = - 2 \zeta_3$ and the computation of  \cite{Bork:2010wf}, where the result $\tilde{f}_1^{(2)} = - 14 \,  \zeta_3$ was found. } 
At this stage, however, we cannot make any prediction for $f^{(2)}_2$ and $C^{(2)}$. In the following we will set $f^{(2)}_2 = -2 \zeta_4$ and $C^{(2)}=0$ so that
\beq
\label{eq:f2}
f^{(2)}= -2 \zeta_2 -2\zeta_3\epsilon -2 \zeta_4\epsilon^2 \ .
\eeq
In this way $f^{(2)}$ matches a closely related quantity appearing in the definition of finite remainders of MHV amplitudes in $\cN=4$ SYM \cite{Bern:2008ap, Drummond:2008aq}
and form factors with $k=2$ \cite{Brandhuber:2012vm}. In order to fix $f^{(2)}_2$ and $C^{(2)}$ individually we would have to calculate also $\F^{{\rm MHV}(2)}_{3, 4}$ and impose that in a collinear limit, where two adjacent momenta $p_i$, $p_{i+1}$ become parallel, the four-point remainder morphs smoothly into the three-point remainder,  
$\cR_{3,4}^{(2)} \to \cR_{3,3}^{(2)}$, without any additional constant. We briefly note here that collinear (and soft) limits
of minimal form factors exhibit novel subtleties compared to amplitudes, and we defer a detailed discussion to Section \ref{sec:discussion}.

Finally, we notice that the $n$-point remainder function  depends on $3n-7$ simple ratios of Mandelstam variables. For $n=3$, we will choose the following variables:
\begin{align}
\label{eq:uvw}
u\ =\ \frac{s_{12}}{q^2},\quad v\ =\ \frac{s_{23}}{q^2},\quad w\ =\ \frac{s_{31}}{q^2},\quad\qquad u+v+w=1\ .
\end{align}

\subsection{The three-point remainder:  from symbols to simple \\ functions}
\label{sec:symbol3}

In the previous two sections we derived the three-point, two-loop form factor and defined its corresponding remainder function. Using the results for the integral functions given in 
\cite{Gehrmann:1999as,Gehrmann:2000zt} we find that the remainder is   a complicated sum of functions of homogeneous degree of transcendentality equal to four which include Goncharov polylogarithms. The expression is rather lengthy and we refrain from presenting it here. However, past experience \cite{Goncharov:2010jf} suggests that one can do much better by studying the symbol of the function. Indeed, this is the strategy we will follow, and at the end we will be able to present a greatly simplified result. 

We find that the symbol of the remainder function is given by the following, strikingly simple expression:
\beq
\label{eq:symbol}
\mathcal{S}_{3,3}^{(2)}(u,v,w)\ =\ 
u \otimes v \otimes  \left[ \frac{u}{w}\otimes_S \frac{v}{w} \right] + 
{1 \over 2}  u \otimes {u \over   (1-u)^3 }  \otimes \frac{v}{w}\otimes
   \frac{v}{w} \, \,   + \,\, \text{perms}\, (u,v,w)\ ,
  \eeq
where $\otimes_S$ in the expression above stands for the symmetrised tensor product
\begin{equation}
x \otimes_S y\,:=\, x \otimes y + y\otimes x \ .
\end{equation}
Before reconstructing the remainder from its symbol \eqref{eq:symbol},  we wish to describe a few general properties
of this remainder and compare them with the properties of symbols of other known remainders of amplitudes and form factors.

{\bf 1.} All entries are taken from the list $\{u,v,w,1-u,1-v,1-w\}$. This is the same list found
for the three-point, two-loop form factor remainder of $\Tr(\phi^2)$ \cite{Brandhuber:2012vm} but does not include the square-root arguments $y_u,y_v,y_w$ present in the case of the two-loop and three-loop, six-point amplitudes in $\cN=4$ SYM 
\cite{Goncharov:2010jf, Dixon:2011pw, Dixon:2011nj}.

{\bf 2.}
The first entries of the symbol describe the locations of discontinuities of the remainder and from unitarity we know that cuts should originate at $P_J^2=0$ or $P_J^2=\infty$, where $P_J^2$ are appropriate kinematic invariants -- in our case $s_{12},s_{23},s_{31}$ and $q^2$.
Hence, the first entry condition \cite{Gaiotto:2011dt} implies in our case that the first entries must be taken from the list $\{u,v,w\}$,  which is obviously the case for \eqref{eq:symbol}.

{\bf 3.} 
In the literature on amplitudes various other conditions on {\it e.g.}  the second and final entries were put forward. However the symbol \eqref{eq:symbol} does  not follow the
pattern observed for two-loop amplitudes or two-loop form factors of $\Tr(\phi^2)$.
For the second entries we observe that if the first entry is $u$ then the second entry is taken from the list $\{u,v,w,1-u \}$,  while the last entry is always an element of the list
$\{u/v, v/w, w/u\}$. We note that the same entry conditions are true for the building blocks of the $k$-point, two-loop form factors of $\Tr( \phi^k)$, which we will discuss in the next sections.
A possible reason why these entry conditions deviate from those of amplitudes and form factors of $\Tr(\phi^2)$ is related to the fact that the form factors we study here have unconventional factorisation properties in collinear and soft limits,  as discussed in Section \ref{sec:discussion}.

We now move on to reconstructing  the remainder from its symbol \eqref{eq:symbol}.
In our case the original expression of the remainder contains many Goncharov polylogarithms, as well as classical (poly)logarithms. However, there is a sharp criterion proposed by Goncharov \cite{gonch, Goncharov:2010jf} that allows us to test if a function of transcendentality four can be re-written in terms of classical polylogarithms $\text{Li}_k$ with $k\leq 4$ only. This criterion is expressed at the level of the symbol as
\beq
\label{eq:gonch-cond}
\cS_{abcd} - \cS_{bacd} - \cS_{abdc}+ \cS_{badc} - (a \leftrightarrow c\, ,\,  b \leftrightarrow d)\ = \ 0
\ .
\eeq
This condition can be rephrased as \cite{Golden:2014xqa}
\beq 
\label{eq-gonch-cond2}
\delta (\cS )\Big|_{\Lambda^2 B_2} = 0 \, ,
\eeq
where the $\Lambda^2 B_2$ component of a symbol (coproduct) is defined as \cite{Golden:2014xqa}%
\footnote{To be more precise we should note that
$\Lambda^2 B_2$ is defined in \cite{Golden:2014xqa} as a particular component of the coproduct 
$\delta$ of a function, but in this paper we will work always at the level of the symbol of the function. The same comment applies to the ${B_3 \otimes \mathbb{C}}$ component of the coproduct introduced later.}
\beq \label{eq:delta-B2}
\delta(a \otimes b \otimes c \otimes d)\Big|_{\Lambda^2 B_2} := \left( a \wedge b \right) \wedge \left( c \wedge d \right) 
\ , 
\eeq \black
and $\wedge$ stands for the anti-symmetrised tensor product
\begin{equation}
x \wedge y\,:=\, x \otimes y - y\otimes x \ .
\end{equation}
Interestingly, our symbol $\mathcal{S}_{3,3}^{(2)}(u,v,w)$ \eqref{eq:symbol} satisfies the constraint \eqref{eq:gonch-cond}, or equivalently \eqref{eq-gonch-cond2}.

A strategy to accomplish this goal was outlined in \cite{Goncharov:2010jf} and starts
by investigating the symmetry properties of the symbol under pairwise (anti)symmetrisation of the entries. In this fashion  one can decompose the symbol into four terms,
\beq
\cS_{3,3}^{(2)}(u,v,w)=\text{A}\otimes\text{A}+\text{S}\otimes\text{A}+\text{A}\otimes\text{S} +\text{S}\otimes\text{S}\ ,
\eeq
where {\it e.g.}~$\text{S}\otimes\text{A}$ means symmetrisation of the first two entries and antisymmetrisation of the last two entries.
Next one scans the symmetry properties of the functions that may appear in the answer, as shown in Table \ref{tab:symmetries} (taken from \cite{Goncharov:2010jf}).
\begin{table}[h]
\centering
\begin{tabular}{|c|c|c|c|c|}
\hline
Function & $\text{A}\otimes\text{A}$ & $\text{S}\otimes\text{A}$ & $\text{A}\otimes\text{S}$ & $\text{S}\otimes\text{S}$ \\
\hline
$\text{Li}_4(z_1)$ & $\times$ & $\times$ & \checkmark & \checkmark \\
\hline
$\text{Li}_3(z_1)\, \log(z_2)$ & $\times$ & $\times$ & \checkmark & \checkmark \\
\hline 
$\text{Li}_2(z_1)\, \text{Li}_2(z_2)$ & \checkmark & \checkmark & \checkmark & \checkmark \\
\hline
$\text{Li}_2(z_1) \, \log(z_2)\, \log(z_3) $ & $\times$ & \checkmark & \checkmark & \checkmark \\
\hline
$\log(z_1)\, \log(z_2)\, \log(z_3)\, \log(z_4) $ & $\times$ & $\times$ & $\times$ & \checkmark \\
\hline
\end{tabular}
\caption{\it Symmetry properties of the symbol of transcendentality four functions.}
\label{tab:symmetries}
\end{table}

Remarkably, we find that our symbol \eqref{eq:symbol}  satisfies even more stringent constraints than \eqref{eq:gonch-cond}, namely
its $\text{A}\otimes\text{A}$ and $\text{S}\otimes\text{A}$ components both vanish. Inspecting Table \ref{tab:symmetries}, we see that $\text{Li}_2$ functions are absent and
only the following functions
\beq
\label{eq:listoffunctions}
\big\{\,\text{Li}_4(z_1),\;\text{Li}_3(z_1) \log(z_2),\;\log(z_1)\log(z_2)\log(z_3)\log(z_4) \, \big\}
\eeq
can appear in the answer.
Goncharov's theorem does not predict what the possible arguments of these functions should be.  
We find that with the following list of arguments 
\beq
\label{eq:listofarguments}
\left\{u,v,w,1-u,1-v,1-w,-\frac{u}{v},-\frac{u}{w},-\frac{v}{u},-\frac{v}{w},-\frac{w
   }{u},-\frac{w}{v},-\frac{u v}{w},-\frac{u w}{v},-\frac{v w}{u}\right\}\ ,
\eeq 
we can construct an ansatz for the result which reproduces the symbol of the remainder
\eqref{eq:symbol}.

Following this procedure we find that the result for the integrated symbol is a remarkably compact two-line function: 
\begin{align}
\label{eq:integratedsymbol}
\begin{split}
\cS_{3,3}^{(2)\,\text{Int}}\ &=\
  \frac{3}{4} \, \text{Li}_4\left(-\frac{u v}{w}\right)-\frac{3}{2} \, \text{Li}_4(u)
-\frac{3}{2} \, \log(w) \text{Li}_3 \left(-\frac{u}{v} \right)\\
&
+\,\frac{ {\log}^2(u)}{32} \Big[ {\log}^2(u) +2 \log^2(v)-4  \log(v) \log(w) \Big]
+ {\rm perms}\, (u,v,w) \ .
   \end{split}
\end{align}
The appearance of the  combination of ${\rm Li}_4$ functions  in \eqref{eq:integratedsymbol} with their particular arguments can in fact be inferred by analysing  the $B_3 \otimes \mathbb{C}$ component of the coproduct $\delta$ \cite{Golden:2014xqa}. At the level of the symbol, this component projects out terms which can be written as symbols of products of functions of lower transcendentality. It is defined as 
\begin{align} 
\label{eq:B3C}
\begin{split}
\delta(a \otimes b \otimes c \otimes d)\Big|_{ B_3 \otimes \mathbb{C} } := 
\left( (a \wedge b) \otimes c  -  (b \wedge c) \otimes a  \right) \otimes d
-
\left( (b \wedge c) \otimes d  -  (c \wedge d) \otimes b  \right) \otimes a  \, , \\[6pt]
\end{split}
\end{align}
which is identical to the definition of the projection operator $\rho$ introduced in 
\cite{Duhr:2011zq}.
For our three-point remainder, which consists only of classical polylogarithms, this implies that we  project on the ${\rm Li}_4$ part of the 
remainder. We find that 
\beq  \label{eq: B3R3}
\delta ( \mathcal{S}_{3,3}^{(2)}(u,v,w) )\Big|_{B_3 \otimes \mathbb{C}} 
= 
-{3 \over 2} \{ u \}_3 \otimes u + { 3 \over 4 } \left\{ - {u v \over w} \right\}_3 \otimes { u v \over w} \, ,
\eeq
where we have introduced the  shorthand notation: 
\beq
\{ x \}_k :=  \{ x \}_2 \otimes^{k-2} x \, , \quad { \rm with} \quad \{ x \}_2 := - (1-x) \wedge x \, . 
\eeq
Noting that 
\beq
\delta ( {\rm Li}_4 (x)  )\Big|_{B_3 \otimes \mathbb{C}} \ = \ \{ x\}_4 \, = \, \{ x\}_3 \otimes x
\ , 
\eeq
we immediately infer from \eqref{eq: B3R3} that the remainder   $\mathcal{R}^{(2)}_{3,3}$
is given by  
$-{3 \over 2} {\rm Li}_4 ( u  ) + \frac{3}{4} {\rm Li}_4 \left( - \dfrac{uv}{w}  \right) $, modulo products of lower transcendentality functions.

\begin{figure}[htb]
\centering
\includegraphics[width=1\textwidth]{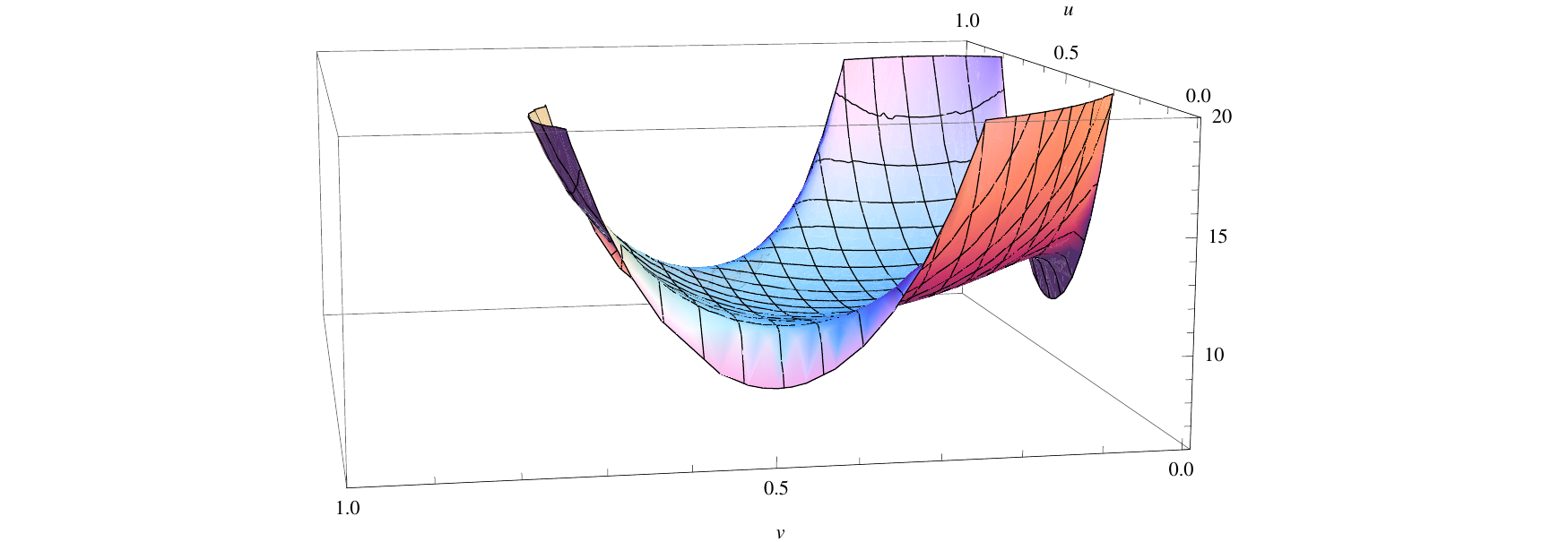}
\caption{\it Plot of the remainder function $\cR_{3,3}^{(2)} (u,v,1-u-v)$, where $u$ and $v$ live in a triangular region bounded by $u=0.01$, $v=0.01$ and $u+v=0.99$. As we approach the edges, for instance $u=0$, the remainder diverges as $\log^2 u$, as explained in the text.}
\label{fig:plotR3}
\end{figure}

The function \eqref{eq:integratedsymbol} is not yet the full remainder because the symbol is blind to transcendentality four functions containing powers of $\pi$ or $\zeta_i$. 
In order to fix these ambiguities, we subtract \eqref{eq:integratedsymbol} from the full remainder function and inspect what is left over. These so-called ``beyond the symbol'' terms 
 are a linear combination of terms of the form $\pi^2 \log x \log y$, $\pi^2 \, \text{Li}_2(x)$, $\zeta_3 \log x$ and $\zeta_4$ and their coefficients can be determined numerically.
In performing the numerical comparison with the original remainder we have used the \texttt{GiNaC} software \cite{DBLP:journals/corr/cs-SC-0004015}. We find the following  result  for the beyond the symbol terms: 
\begin{align}
\label{eq:beyondthesymbol}
\cR_{3,3}^{(2)  \text{bts}} =\   {\zeta_2 \over 8}  \log(u)\, \Big[ 5 \log(u)
    -2 \log (v) \, \Big] + \frac{\zeta_3}{2}  \log (u) +\frac{7}{16}\zeta_4  + {\rm perms}\, (u,v,w)  \ .
\end{align}
Summarising, the final result for the remainder function  $\cR_{3,3}^{(2)}$ is the sum of \eqref{eq:integratedsymbol} and \eqref{eq:beyondthesymbol}, 
\begin{align}
\label{eq:remainderT3}
\begin{split}
\cR_{3,3}^{(2)} \ := \ & 
 -\frac{3}{2}\, \text{Li}_4(u)+\frac{3}{4}\,\text{Li}_4\left(-\frac{u v}{w}\right)  
-\frac{3}{2}\log(w) \, \text{Li}_3 \left(-\frac{u}{v} \right)   
+ \frac{ 1}{16}  {\log}^2(u)\log^2(v)
\\
&
+ {\log^2 (u) \over 32} \Big[ \log^2 (u) - 4 \log(v) \log(w) \Big]  
+  {\zeta_2 \over 8 }\log(u) [  5\log(u)- 2\log (v) ]
 \\
&
+  {\zeta_3 \over 2} \log(u) + \frac{7}{16}\, \zeta_4  + {\rm perms}\, (u,v,w) \ .
\end{split}
\end{align}
We plot the remainder function $\cR_{3,3}^{(2)} (u,v,1-u-v)$ in Figure \ref{fig:plotR3}.

One important feature which stands out is that the remainder  blows up at the boundaries of the Euclidean kinematic region $u=0$, $v=0$ and $u+v=1$. We need to distinguish here two types of limits: 

{\bf 1.} The situation where we approach a generic point on one of the three edges corresponds to a collinear limit. For instance,  taking $u\to 0$ (and $v+w\to 1$) is equivalent to the collinear limit $p_1 \, ||\, p_2$. In this situation
the remainder diverges as $\log^2 (u)$. The derivation of this result can be found in  Section \ref{sec:discussion}. 

{\bf 2.} The case where we approach one of the corners, for instance $u=w=0$, corresponds to the soft limit $p_1 \to 0$. As will be discussed in Section \ref{sec:discussion}, this soft limit can be parametrised as 
$ u=x \, \delta$, $v = 1 - \delta$, $w=y\,\delta$ with $x+y=1$ and $\delta \to 0$ and the remainder diverges as $(1/4) \log^4 (\delta)$, explaining the spikes in  Figure \ref{fig:plotR3} in the positive vertical direction.

This behaviour might appear  unexpected for  remainder functions, which usually have smooth  collinear and soft limits. However one has to appreciate that here we are considering a special form factor, with the minimal number of external legs. 
Hence we cannot extrapolate the usual intuition about factorisation since there is no form factor with fewer legs  this minimal form factor could factorise on, as we discuss in more
detail in Section \ref{sec:discussion}.

\section{The two-loop remainder function for all $k>3$}
\label{sec:allk}

Having obtained and described in detail the three-point remainder  $\cR_{3,3}^{(2)}$ of the form factor of the operator ${\rm Tr }[ ( \phi^{++})^3]$ at two loops,  we now move on to study the $k$-point form factors  $\F_{k,k}^{(2)}$ of  ${\rm Tr }[ ( \phi^{++})^k]$ for arbitrary $k>3$.

\subsection{The $k$-point minimal form factors  from cuts}

The study of the cuts of these form factors proceeds in an almost identical way compared to the $k=3$ case, with one important exception, namely the appearance of a new integral function which is the product of two one-loop triangle functions. Specifically, our result for the  
minimal form factor of ${\rm Tr }[ ( \phi^{++})^k]$ for $k>3$  at two loops is given by the following simple  extension of that of $k=3$,
\beq \label{eq:basisTk}
\cF^{(2)}_{k,k}\,=\, \sum_{i=1}^n\Big[ I_1(i) + I_2(i) + I_3(i) +I_4(i) - I_5(i) +  {1\over 2} \sum_{j=i+2}^{i-2} I_6(i, j) \Big] \, ,
\eeq
where the integral basis is the same as \eqref{eq:basis} augmented by one new integral, namely $I_6$:
\beq
\label{eq:basis2}
\vcenter{\hbox{\includegraphics[width=.92\linewidth]{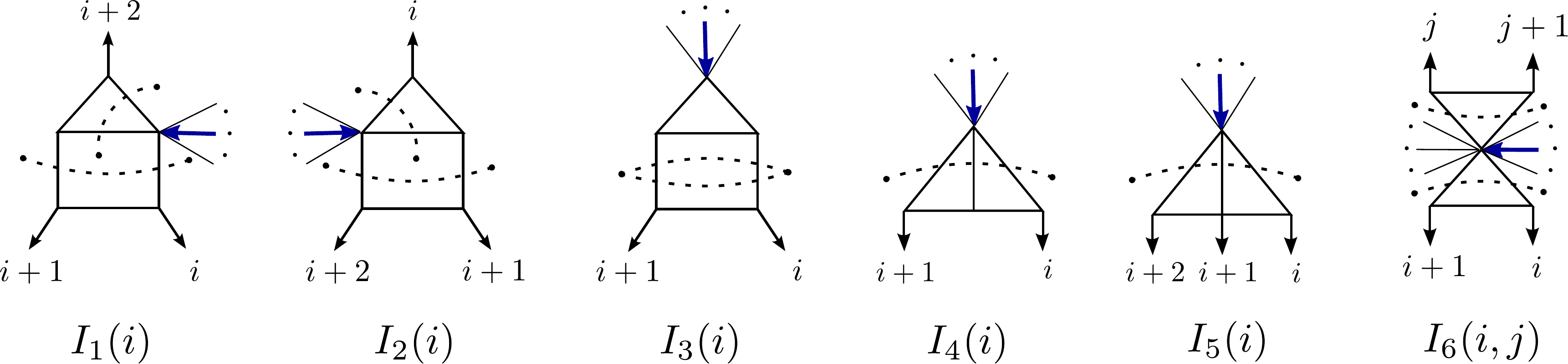}}} \, .
\eeq
The factor of $1/2$ in front of $I_6$ is present in order to remove double counting.%
\footnote{A similar but not identical result for the same quantities was presented in \cite{Bork:2010wf}. As in the three-point case discussed earlier, our result differs from theirs by the absence of the function $G_3$ appearing in Eqn.~(4.44) of \cite{Bork:2010wf}. }

Next, we  comment that the appearance of the extra integral function $I_6$ can be inferred easily from two-particle cuts, specifically   by attaching \eqref{eq:attach} (with $2$ and $3$ replaced by $i$ and $i+1$) to the following one-loop triangle integral, 
\beq 
\label{uuuuu}
\vcenter{\hbox{\includegraphics[scale=0.42]{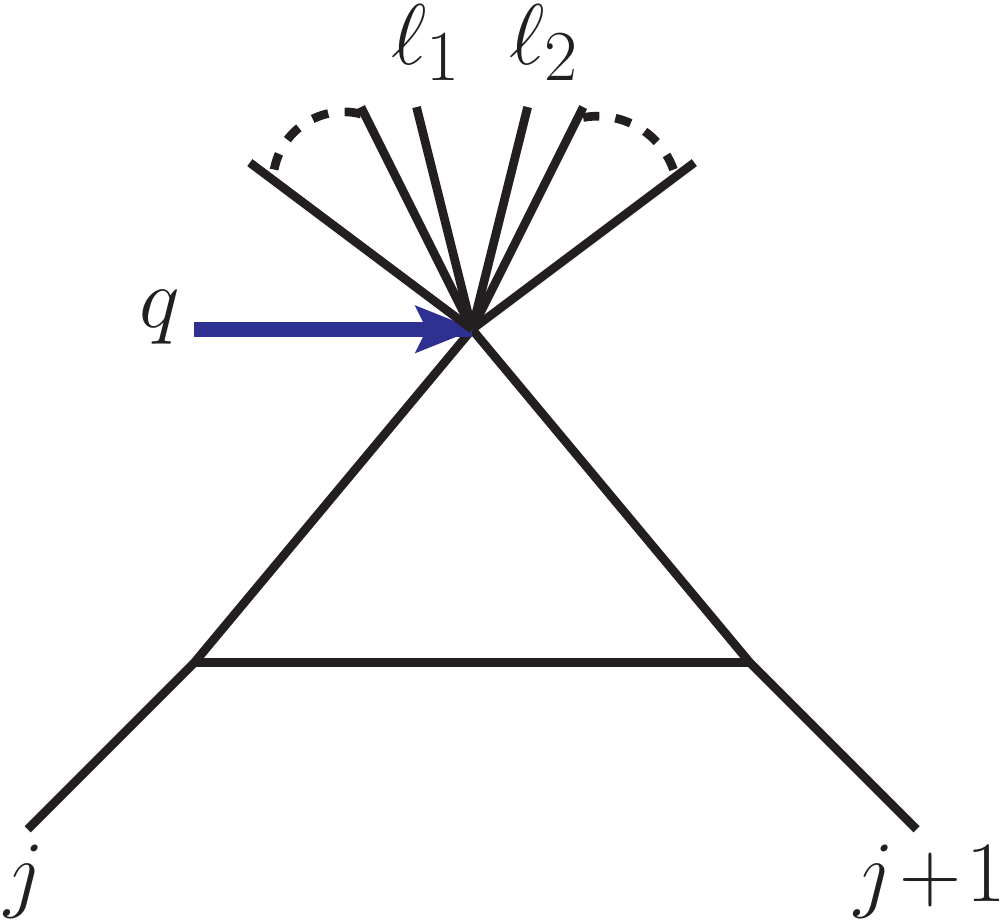}}}\, 
\eeq
Clearly, \eqref{uuuuu} is  present only when   $k>3$. 
With this additional term, \eqref{eq:basis2} has  all the correct two-particle cuts.

Let us now discuss how the  following triple cuts might get altered when compared to the $k=3$ case studied earlier.
\begin{figure}[h]
\centering
\includegraphics[width=0.85\linewidth]{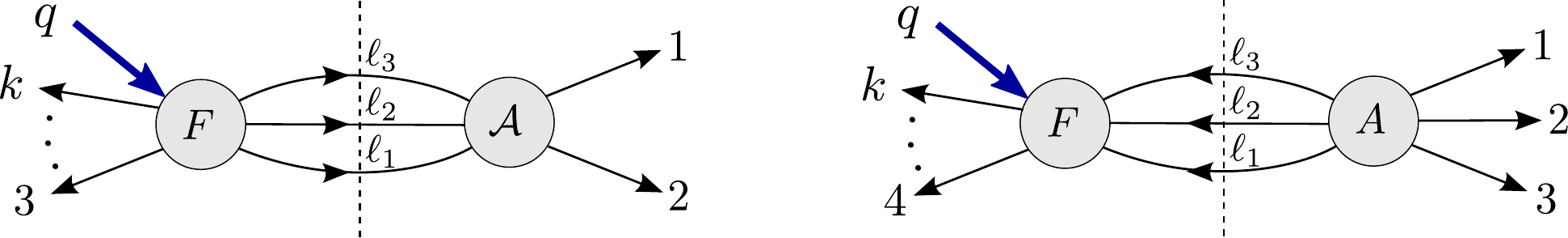}
\caption{\it Triple cuts employed in the derivation of  $\F_{k,k}^{(2)}$.}
\label{eq:tri-cut-k}
\end{figure}
To begin with, we note that $I_6$ does not contribute to any triple cut. Since the remaining integrals in  \eqref{eq:basis2} are identical to those contributing to  $\F_{3,3}^{(2)}$, we  only need to confirm that the results of the above triple cuts are the  same as those of $\F_{3,3}^{(2)}$. 

This agreement is immediate for the diagram on the right-hand side of  Figure \ref{eq:tri-cut-k} since the form factor appearing there is simply $1$. For the diagram on the left-hand side, a simple way to show this is to note that the tree-level form factors which  enter  the cut are actually identical for $k=3$ and for $k>3$. They are $(k+1)$-point NMHV form factors with one negative-helicity  gluon, $g^-$, or two fermions, $\bar{\psi}$, which indeed take the same form for any $k$, namely
\begin{align} 
\label{fff}
 & F_{k,k+1}^{\rm NMHV}(\phi_1, \ldots, \phi_{i-1}, g^-_i, \phi_{i+1}, \ldots, \phi_{k+1} )
= {[i\!-\! 1 \, i\!+\! 1] \over [i\!-\! 1 \, i] [i \, i\!+\! 1 ]} \, , \nonumber \\
& F_{k,k+1}^{\rm NMHV}(\phi_1, \ldots, \phi_{i-1}, \bar{\psi}_{i}, \bar{\psi}_{i+1}, \phi_{i+2}, \ldots, \phi_{k+1} )
= {1 \over  [i \, i\!+\! 1 ]} \, ,  \\ \nonumber
& F_{k,k+1}^{\rm NMHV}(\phi_1, \ldots, \bar{\psi}_{i-1}, \phi_{i}, \bar{\psi}_{i+1}, {\phi}_{i+2}, \ldots, \phi_{k+1} )
= 0 \, .
\end{align}
The above results \eqref{fff} can be obtained simply by taking the conjugate of the $(k+1)$-point MHV form factors of ${\rm Tr}(\phi^k)$.

In conclusion, compared to the case  $k=3$, the only difference in the result is that now we need to include the double-triangle integrals $I_6(i,j)$.

\subsection{The symbol of the $k$-point remainder}

In this section we construct the two-loop remainder function and its symbol for the case of general $k$. The remainder is defined in \eqref{eq:remainder}, where now $k=n>3$. The ingredients of this formula are the one-loop minimal form factor defined in \eqref{eq:1-loop-MHV-all-k}, and the two-loop form factor derived earlier in this section. A few  comments are in order.

{\bf 1.} We find that the cancellation of the infrared poles in $\epsilon$ proceeds exactly as in the three-point case, and as a result the remainder function is defined with the same universal function $f^{(2)} (\eps)$ defined in \eqref{eq:f2}.

{\bf 2.} As noticed earlier, the two-loop form factor contains an extra integral topology $I_6$ if $n\!=\!k\!>\!3$. This topology is exactly cancelled by the cross terms coming from the square of the one-loop form factor appearing in the definition of the remainder. There is an important consequence of this cancellation, namely all the remaining integral topologies  contributing to the remainder depend only on either triplets of adjacent momenta $p_i$, $p_{i+1}$ and $p_{i+2}$  ($I_1$, $I_2$,  and $I_5$) or pairs of adjacent momenta ($I_3$ and $I_4$).  As a result the remainder function can be written as a cyclic sum over universal sub-remainders which depend on three momenta, 
\begin{align}
\label{eq:3part-bb}
\cR_{k,k}^{(2)}\,=\, \sum_{i=1}^k r^{(2)}(u_i,v_i,w_i) \ ,
\end{align}
as we will show in detail below. Here the parameters $u_i, v_i, w_i$ are generalisations of the $u, v, w$ ratios of the $k=3$ case, and are defined  as
\begin{align}
\label{eq:uvw2}
u_i\,=\,\frac{u_{i\,i+1}}{u_{i\,i+1 \, i+2}}\, ,\quad
& v_i\,=\,\frac{u_{i+1\,i+2}}{u_{i\,i+1 \, i+2}}\, ,\quad
w_i\,=\,\frac{u_{i+2\,i}}{u_{i\,i+1 \, i+2}} \ ,
\end{align}
with 
\beq
u_{i\,i+1 \, i+2} := u_{i\,i+1}+u_{i+1\,i+2}+u_{i+2\,i} \, .
\eeq 
Note that we have  defined $u_{ij}:=s_{ij}/q^2$, and  $u_i+v_i+w_i = 1$.
For notational simplicity we will in the following replace $r^{(2)}(u_i,v_i,w_i)$ by
$r^{(2)}_{i}$. We should stress at this point that these are the basic building blocks of
$\cR_{k,k}^{(2)}$ and do not depend on the value of $k$.

Using the explicit expressions of the integral functions $I_1, \ldots, I_6$ we have computed the remainder function  in terms of multiple polylogarithms. As in the three-point case, this expression is quite lengthy and we will only present it after simplifying it using its symbol. 

Again it turns out that the symbol is extremely simple. As anticipated above, it is written as a sum of  building blocks which depend on $u_i$, $v_i$ and $w_i$: 
\begin{align}
\cS_{k,k}^{(2)}\,=\, \sum_{i=1}^k s^{(2)}(u_i,v_i,w_i) := \sum_{i=1}^k s^{(2)}_i \ ,
\end{align}
where
\begin{align}  \nonumber
s^{(2)}_i
\ &= \
 u_i\otimes (1-u_i)\otimes \left[ \frac{u_i-1}{u_i} \otimes \frac{v_i}{w_i}+ \frac{v_i}{w_i}\otimes \frac{w_i^2}{u_i v_i} \right] \\ 
& +u_i\otimes u_i\otimes {1-u_i \over v_i} \otimes \frac{w_i}{v_i}  
 +u_i\otimes v_i\otimes  \left[ \frac{v_i}{w_i} \otimes_S \frac{u_i}{w_i} \right]
 + (u_i \leftrightarrow v_i ) \, .
\end{align}
As was done earlier for the $k=3$ case, it is useful to study the coproduct of the remainder function. A key difference is that, unlike  the case of the form factor of $\T_3$, the symbol $s^{(2)}_i$ does not obey Goncharov's condition \eqref{eq:gonch-cond}. Instead  we find that the corresponding component of the coproduct is
\beq 
 \label{eq:BkB2B2}
\delta ( s^{(2)}_i )\Big|_{\Lambda^2 B_2} \ = \ 
\left\{- { w_i \over v_i } \right\}_2 \wedge \left\{ u_i \right\}_2 \ + \ ( u_i \leftrightarrow v_i )
\ . 
\eeq 
We also quote its  $B_3 \otimes \mathbb{C} $ component, given by 
\begin{align}  \label{eq:BkB3C}
\delta ( s^{(2)}_i )\Big|_{B_3 \otimes \mathbb{C}} 
&= 
\{ 1- u_i  \}_3 \otimes {  w_i \over v_i } \ + \ \{ u_i \}_3 \otimes { u_i \over w_i}
\ + \ \left\{ - {w_i  \over u_i v_i} \right\}_3 \otimes { w_i \over u_i} 
\cr
& + \ \left\{ -{ w_i \over v_i} \right\}_3  \otimes { v_i \over u_i w^2_i} 
\ - \  \left\{{ v_i \over 1-u_i } \right\}_3 \otimes u_i
\ + \  \left( u_i \leftrightarrow v_i \right)
\ .
\end{align}
Because of the non-vanishing of the  ${\Lambda^2 B_2}$ component \eqref{eq:BkB2B2}, $s^{(2)}_i$ cannot be integrated to purely classical polylogarithms. However, it is not difficult to recognise what multiple polylogarithms can give rise  to \eqref{eq:BkB2B2}. For instance $ {\rm Li}_{1,3}\Big(u_i, -\dfrac{w_i}{u_i v_i}\Big) \ + \ ( u_i \leftrightarrow v_i )$, or the cluster algebra inspired  function $L_{2,2}\Big(u_i, -\dfrac{w_i}{v_i}\Big) \ + \ ( u_i \leftrightarrow v_i ) $ defined in \cite{Golden:2014xqa} can do the job.

In the present case,  it turns out to be more convenient to consider the following combination of Goncharov polylogarithms, as we will explain momentarily,%
\footnote{These Goncharov polylogarithms already appear in the explicit expressions of the integrals $I_1(i)$ and $I_2(i)$ belonging to the basis \eqref{eq:basis2}.}
\begin{align}
\label{eq:extragonch}
\begin{split}
r_{{\rm nc},i}^{(2)}  \, &:=\,-  G\left(\left\{1-u_i,1-u_i,1,0\right\},v_i\right)\ - \ ( u_i \leftrightarrow v_i ) \, , 
\end{split}
\end{align}
where the symbol of $G_{v}\,:=\, G\left(\left\{1-u,1-u,1,0\right\},v\right)$ is given by
\begin{align}
\begin{split}
\cS[G_v]\ = \ &   
v \otimes w \otimes \big[ w \otimes_S u -  u \otimes u \big]
+ v \otimes(1-v) \otimes {u \over w} \otimes {  u \over w } -v \otimes (1-u) \otimes (1-u) \otimes u   
\\
&
 + {w \over v(1-u)} \otimes  (1-u) \otimes u \otimes {w \over u}
 + {  v(1-u) \over w } \otimes \left[  (1-u) \otimes_S {  1-u \over w }  \right] \otimes u
\ .
\end{split}
\end{align}
As for the functions ${\rm Li}_{1,3}$ and $L_{2,2}$ mentioned previously, the $\Lambda^2 B_2$ component of the coproduct of  
$\cS[r_{{\rm nc},i}^{(2)}]$ is  equal to \eqref{eq:BkB2B2}. Hence we can decompose the symbol of the remainder  $s^{(2)}_i$  into a non-classical and a classical contribution:  
\beqa
s^{(2)}_i\, =\, s_{{\rm nc},i}^{(2)}  +  s_{{\rm cl},i}^{(2)}\ ,
\eeqa
where $s_{{\rm nc},i}^{(2)}$ is the symbol of $r_{{\rm nc},i}^{(2)}$. Hence  $s_{{\rm cl},i}^{(2)}$  has now a vanishing $\Lambda^2 B_2$ component, or equivalently satisfies Goncharov's condition  \eqref{eq-gonch-cond2}, and thus can be rewritten in terms of classical polylogarithms only.  

We now move on to determining the  classical part of the remainder.  In order to do so, it is convenient to  first examine
the  $B_3 \otimes \mathbb{C}$ component of the non-classical remainder $r_{{\rm nc},i}^{(2)} $. It is given by 
\begin{align}
\begin{split}
\delta( s_{{\rm nc},i}^{(2)} )\Big|_{B_3 \otimes \mathbb{C}} 
&=  \{ 1- u_i  \}_3 \otimes { u_i w_i  \over (1- u_i)^2  v_i } \ + \ \{ u_i \}_3 \otimes { u_i \over w_i (1-u_i) }
\ + \ \left\{ - {w_i  \over u_i v_i} \right\}_3 \otimes {w_i \over u_i } 
\cr
& + \ \left\{ -{ w_i \over v_i} \right\}_3  \otimes {v_i \over u_i w^2_i} 
\ - \  \left\{{ v_i \over 1-u_i  } \right\}_3 \otimes u_i
\ + \ 
\left( u_i \leftrightarrow v_i  \right) 
\ .
\end{split}
\end{align}
This is a somewhat complicated expression, however the particular  choice of $r_{{\rm nc},i}^{(2)}$ we made  in \eqref{eq:extragonch} is such that 
the  $B_3 \otimes \mathbb{C}$ component of the coproduct of $r_{{\rm cl},i}^{(2)}$ turns out to be very simple -- in fact this was the motivation behind choosing our particular form of $r_{{\rm nc},i}^{(2)}$. Furthermore, $r_{{\rm nc},i}^{(2)}$ does not develop any singularity  in the soft or collinear limits (this is shown explicitly in Section  \ref{mff},  see  \eqref{cp}). 
For the $B_3 \otimes \mathbb{C}$ component of the classical remainder $r_{{\rm cl},i}^{(2)} $
we find on the other hand
\beq 
\delta\big(s_{{\rm cl},i}^{(2)}\big)\Big|_{ B_3 \otimes \mathbb{C}} 
\ =\
\{1- u_i \}_3 \otimes { (1- u_i )^2 \over u_i } + \{ u_i \}_3 \otimes (1- u_i) 
+  
\left( u_i \leftrightarrow v_i  \right) \, .
\eeq
By applying the  identity $\{ 1- 1/u \}_3 =  - \{ 1- u \}_3 - \{ u \}_3 $, this expression can be recast as 
\beq
\delta\big(s_{{\rm cl},i}^{(2)}\big)\Big|_{ B_3 \otimes \mathbb{C}} 
\ =\
\{1- u_i \}_3 \otimes (1- u_i)  + 
\{ u_i \}_3 \otimes u_i
- 
\{1- {1 \over u_i} \}_3 \otimes \Big(1- {1 \over u_i} \Big) +  
\left( u_i \leftrightarrow v_i  \right)\, . 
\eeq
From the above result, we see immediately that the classical part of the remainder  $r_{{\rm cl},i}^{(2)}$  is  given by  ${\rm Li}_4(1-u_i) + {\rm Li}_4(u_i) - {\rm Li}_4(1-1/u_i)+  
\left( u_i \leftrightarrow v_i  \right)$ modulo  products of functions of lower degree of transcendentality, which  can be fixed by following the same strategy as in the  $k=3$ case. 
Doing so, we find that the classical part of the remainder is: 
\begin{align}
\label{eq:intmodsymbolk}
\begin{split}
r_{{\rm cl},i}^{(2)} =\, &    \text{Li}_4(1-u_i)+\text{Li}_4(u_i)-\text{Li}_4\left(\frac{u_i - 1}{u_i}\right)
+  \log \left( {1-u_i \over w_i }\right) 
\left[  \text{Li}_3\left(\frac{u_i - 1}{u_i}\right) - \text{Li}_3\left(1-u_i\right) \right] \\
 +\, & \log \left(u_i\right) \left[\text{Li}_3\left(\frac{v_i}{1-u_i}\right)+\text{Li}_3\left(-\frac{w_i}{v_i}\right) + \text{Li}_3\left(\frac{v_i-1}{v_i}\right)
 -\frac{1}{3}  \log ^3\left(v_i\right) -\frac{1}{3} \log ^3\left(1-u_i\right)  \right]
 \\   
+\, & \text{Li}_2\left(\frac{u_i-1}{u_i}\right) \text{Li}_2\left(\frac{v_i}{1-u_i}\right)-  \text{Li}_2\left(u_i\right) \left[
   \log \left({1-u_i \over w_i }\right) \log \left(v_i \right) +\frac{1}{2} \log ^2\left( { 1-u_i \over w_i }\right) \right] 
\\
-\, & \frac{1}{24} \log ^4\left(u_i\right)
+\frac{1}{8} \log ^2\left(u_i\right) \log ^2\left(v_i\right)  + \frac{1}{2} \log ^2\left(1-u_i\right) \log \left(u_i\right) \log \left( { w_i \over v_i}\right)\\
+\, & \frac{1}{2} \log \left(1-u_i\right) \log ^2\left(u_i\right) \log \left(v_i\right)
+ \frac{1}{6} \log ^3\left(u_i\right) \log \left(w_i\right) \ +\  (u_i\, \leftrightarrow\, v_i) \ .
\end{split}
\end{align}
The beyond the symbol terms (obtained in the same way as for $\cR^{(2)}_{3,3}$) are
\begin{align}
\label{eq:btsk}
\begin{split}
r_{\text{bts},i} ^{(2)} = \ & \zeta_2 \, \Big[ \log \left(u_i\right) \log \left(1-v_i \over v_i \right)
+ \frac{1}{2}\log ^2\left( { 1-u_i \over w_i }\right) - \frac{1}{2}\log ^2\left(u_i\right)   \Big]
\\
-\, &   \zeta_3 \log (u_i)  - {\zeta_4\over 2} 
+\  (u_i\, \leftrightarrow\, v_i)
 \ .
\end{split}
\end{align}
\renewcommand{\arraystretch}{1.5}
\begin{table}[t]
\centering
\begin{tabular}{|c||c|}
\hline
$  k $ &  Estimated error\hspace{10pt} \\
\hline  $4$ & $\cO(10^{-17})$
\\
\hline  $5$ & $\cO(10^{-14})$ \\
\hline  $6$ & $\cO(10^{-15})$ \\
\hline
\end{tabular}
\caption{\it Numerical checks of the remainders $\cR^{(2)}_{k,k} $ for $k\,=\,4,\,5,\,6$.}
\label{tab:num-checks}
\end{table}
Finally, the two-loop remainder function for general $k$ is given by
\begin{equation}
\label{eq:remainderk} 
\cR^{(2)}_{k,k} \ = \ \sum_{i=1}^k \Big[
r_{{\rm nc},i}^{(2)}  +  r_{{\rm cl},i}^{(2)}+r_{\text{bts},i} ^{(2)} \Big]\ ,
\end{equation}
where $r_{{\rm nc},i}^{(2)}$, $ r_{{\rm cl},i}^{(2)}$ and $r_{\text{bts},i}^{(2)}$, and  are defined  in  \eqref{eq:extragonch},  \eqref{eq:intmodsymbolk} and  \eqref{eq:btsk},  respectively.

We have also checked  our result \eqref{eq:remainderk}  against numerical evaluations of the remainder for several  values of $k$ and sets of kinematical data,  finding excellent agreement (see  Table~\ref{tab:num-checks}).

\section{Collinear and soft limits}
\label{sec:discussion}

In this section we wish to discuss some general properties of the form factors
under soft and collinear limits. This discussion is somewhat beyond the main
line of the paper, but will be relevant for future studies of non-minimal form factors.

When discussing collinear or soft limits it is crucial to distinguish the cases of minimal
and non-minimal form factors. In the latter case, the number of external on-shell particles
is larger than the number of fields in the operator, and the factorisation properties 
are identical to those of amplitudes. 
This follows from a slight generalisation of arguments presented in \cite{Brandhuber:2012vm}
for form factors of $\Tr (\phi^2)$ with three or more external particles, which in turn
are inspired by the original proof for amplitudes given in \cite{Kosower:1999xi}.
For minimal form factors,  which are the main focus of this paper,  the story is more interesting
since they cannot factorise onto form factors with fewer legs. Hence, the argument of \cite{Kosower:1999xi} does not apply and the factorisation
properties deviate dramatically from those of amplitudes.

\subsection{Minimal form factors}
\label{mff}

We begin by looking at minimal form factors, and specifically we wish to study the collinear and soft behaviour  of their remainder functions derived in the previous sections.  

For non-minimal form factors, one can define a properly normalised  $n$-point remainder function%
\footnote{At two-loop level the appropriate normalisation is obtained by introducing the   $n$-independent, transcendentality-four constant  $C^{(2)}$ in the definition \eqref{eq:remainder}.}
such that, under a collinear limit one has
\beq
\label{usual}
\cR_n \to \cR_{n-1}
\ . 
\eeq
Note that \eqref{usual} is the usual behaviour of remainders of loop amplitudes in $\cN=4$ SYM as discussed in \cite{Bern:2008ap, Anastasiou:2009kna}, and confirmed for the case of form factors of $\Tr \, (\phi^2)$ in \cite{Brandhuber:2012vm}.

As already mentioned in Section \ref{sec:symbol3} (see Figure \ref{fig:plotR3}), this is not possible for the case of a minimal remainder function. 
This is caused by the simple fact that tree-level minimal form factors are $1$,  and remain $1$ under collinear/soft limits.  In what follows we will  quantify the failure to obey conventional factorisation. It is worth stressing that this failure only affects finite terms,  while the universality of infrared divergences also extends to the minimal form factors. This
is related to the fact that we were able to define a finite remainder function for minimal form factors \eqref{eq:remainder} in complete analogy with scattering amplitudes in $\cN=4$ SYM and non-minimal form factors of $\Tr (\phi^2)$ in \cite{Brandhuber:2012vm}.

We begin our study with the simplest remainder function,  namely $\cR^{(2)}_{3,3}$ given in  \eqref{eq:remainderT3}. We consider  the collinear limit $p_1\,||\,p_2$, which we parameterise as 
\begin{equation} 
\label{collinear12}
p_1 \rightarrow z P \,, \quad  p_2  \rightarrow (1 - z) P \,, \qquad P^2=0\ .
\end{equation}
In terms of the $u,v,w$ variables, this is equivalent to
\begin{equation}
u \rightarrow 0 \, , \quad  v \rightarrow (1-z) \, ,  \quad w \rightarrow z \, . 
\end{equation} 
In the limit \eqref{collinear12}, an explicit calculation shows that  
\begin{equation}
\cR^{(2)}_{3,3}(u, v, w) \,{\buildrel 1 \parallel 2\over
{\relbar\mskip-1mu\joinrel\longrightarrow}}
\, 
 \sum^{2}_{m=1} \log^m(u) \ C_{3;m}(z)\, , 
\end{equation}
where the coefficients $C_{3;m}(z)$ are given by
\begin{align}
\begin{split}
C_{3;2}(z) &= {1 \over 4} \left[ \log^2\left( {z \over 1-z} \right) - 2 \zeta_2 \right] \, ,\\[5pt]
C_{3;1}(z) &= -C_{3;2}(z) \log \left[ z (1 - z) \right] + 
 {3 \over 2} \left[ {\rm Li}_3\left({ z \over z-1}\right) + {\rm Li}_3 \left({z-1 \over z}\right) \right] - \zeta_3 \, .
\end{split}
\end{align}
Next, we consider  the soft limit $p_1 \rightarrow 0$, where we have to take $z \rightarrow 0$
in addition to $u \rightarrow 0$. 
Equivalently, one can parametrise the soft limit as
\begin{align}
u = x \, \delta\, ,\qquad v = 1-  \delta\, ,\qquad w = (1 -x )\, \delta\, ,
\end{align}
with $\delta \rightarrow 0$. In this  limit we find
\begin{equation}
\cR^{(2)}_{3,3}(u, v, w) 
\ {\buildrel p_1 \to \, 0\over
{\relbar\mskip-1mu\joinrel\longrightarrow}}
\  
 \sum^{4}_{m=1} \log^m( \delta ) \ S_{3;m}(x)\ ,
\end{equation}
where the coefficients $S_{3;m}(x)$ at each order are given by
\begin{align}
\begin{split}
S_{3;4}(x) &\,=\, {1 \over 4} \, ,\\
S_{3;3}(x) &\,=\,  {1 \over 2 } \log \big[ x(1 - x)   \big] \, , \\
S_{3;2}(x) &\,=\,  [S_{3;3}(x)]^2 +  {1 \over 2} \log(1 - x) \log(x) + \zeta_2 \, ,  \\ 
S_{3;1}(x) &\,=\, 2 \Big( S_{3;3}(x) S_{3;2}(x)- [S_{3;3}(x)]^3 - \zeta_3   \Big) \, .
\end{split}
\end{align}
Now we turn our attention to the study of $\cR^{(2)}_{k,k}$ with $k>3$, in particular we will analyse the behaviour of the three-particle building blocks $r^{(2)}_i$ defined in \eqref{eq:3part-bb}. For the collinear limit $p_1 \,||\, p_2$  introduced in \eqref{collinear12}, both $r^{(2)}_i$ and $r^{(2)}_k$ contribute. Here we focus on $r^{(2)}_1$ only, since $r^{(2)}_k$ behaves in a similar  way.

We begin by observing that  $r_{{\rm nc},i}^{(2)}$ is regular as $u_i \rightarrow 0$, specifically
\begin{align}
\label{cp}
\begin{split}
 \lim_{u_i \rightarrow  0 }  r_{{\rm nc},i}^{(2)} =\, & 0- G\left(\left\{1,1,1,0\right\},v_i\right)\\
  = & - {1 \over 6}\log^2(1-v_i)\big[ \log(v_i) \log(1-v_i) 
 + 
3{\rm Li}_2(v_i) \big]
\\
& -\, 
\log(1-v_i)\, {\rm S}_{1,2}(v_i)
+{\rm S}_{1,3}(v_i) \, , 
\end{split}
\end{align}
where $\mathrm{S}_{n,p}(z)$ denotes a Nielsen polylogarithm. On the other hand, if we now consider  the limit $w_i \rightarrow 0$ we observe that   this function develops a $\log^2(w_i)$ singularity. 
This singularity is required in order to cancel an identical and opposite  singularity arising  from $r^{(2)}_{{\rm cl}, i} + r_{\text{bts},i} ^{(2)}$. 
This is expected since $w_i \rightarrow 0 $ corresponds to two non-adjacent legs becoming collinear, which is not a physical singularity.

Setting in the collinear limit
\begin{equation}
u_1 \rightarrow 0 \, , \quad  v_1 \rightarrow (1-z) \, ,  \quad w_1 \rightarrow z \, ,
\end{equation}  
we obtain
\begin{equation}
r^{(2)}_1 
\,{\buildrel 1 \parallel 2\over
{\relbar\mskip-1mu\joinrel\longrightarrow}}
\, 
\sum^{2}_{m=1} \log^m(u_1)\, C^{1 \parallel 2}_{k;m}(z)\ ,
\end{equation}
where
\begin{align}
\begin{split}
C^{1 \parallel 2}_{k;2}(z) &= {1 \over 2} \left( {1\over 2} \, {\log^2(1-z)} +  {\rm Li}_2(z) - \zeta_2 \right)\ ,\\
C^{1 \parallel 2}_{k;1}(z) &= {1 \over 2} \log^2(1-z) \log(z)  -  {1 \over 3} \log^3(1-z) + 2 \, {\rm Li}_3 \left( {z \over z-1} \right) + 
 {\rm Li}_3(1-z) - \zeta_3 \ .
\end{split}
\end{align}
Finally, we consider the soft limit  for $r^{(2)}_1$. Because of the lack of permutation symmetry, $r^{(2)}_1$ behaves differently under the limits $p_1 \rightarrow 0$ and $p_2 \rightarrow 0$. In the limit $p_2 \rightarrow 0$, or equivalently
\beq
u_1 \,=\, x \,  \delta\, ,\quad v_1 \,=\, (1-x)\, \delta\, ,\quad  w_1\,=\, 1- \delta\, , 
\eeq
with $\delta \rightarrow 0$, we have
\begin{equation} \label{soft123}
r^{(2)}_1
\,{\buildrel p_2 \to \, 0 \over
{\relbar\mskip-1mu\joinrel\longrightarrow}}
\, 
 \sum^{4}_{m=1}  \log^m( \delta ) \, S^{p_2}_{k;m}(x)\ ,
\end{equation}
with
\begin{align}
\begin{split}
S^{p_2}_{k;4}(x) &\,=\, {1 \over 4} \, ,\\
S^{p_2}_{k;3}(x) &\,=\, {1 \over 2} \log \big[ x (1-x) \big] \, ,
\cr
S^{p_2}_{k;2}(x) &\,=\,  2 \zeta_2 + [S^{p_2}_{k;3}(x)]^2 + {1 \over 2} \log(x) \log(1-x)  \, , \cr 
S^{p_2}_{k;1}(x) &\,=\, S^{p_2}_{k;3}(x) \Big[ 4 \zeta_2 +\log(x) \log(1-x)  \Big] - 2\zeta_3 \, 
.
\end{split}
\end{align}
For $p_1 \rightarrow 0$, or equivalently
\beq
u_1\, =\, x\, \delta\, ,\quad  v_1\, =\,1 - \delta\, ,\quad w_1\,=\, (1-x)\, \delta\ ,
\eeq 
with $\delta \rightarrow 0$, we have
\begin{equation}
r^{(2)}_1
\,{\buildrel p_1 \to 0\over
{\relbar\mskip-1mu\joinrel\longrightarrow}}
\, 
 \sum^{4}_{m=1} \log^m( \delta ) S^{p_1}_{k;m}(x)\ ,
\end{equation}
where
\begin{align}
S^{p_1}_{k;4}(x) &\,=\,S^{p_1}_{k;3}(x) \,=\, S^{p_1}_{k;2}(x)\,=\,0  \, , \cr 
S^{p_1}_{k;1}(x) &\,=\, \zeta_2 \log \left(  {1 - x \over x }  \right) - \zeta_3 \, .
\end{align}
Note that  $r^{(2)}_1$ is less singular as $p_1 \rightarrow 0$ compared to the previous case where   $p_2 \rightarrow 0$. However, the full remainder is completely symmetric and should behave in the same way for arbitrary $p_i \rightarrow 0$. Indeed it is the building block with external legs $p_{k}, p_1, p_2$, namely $r^{(2)}_k$, that carries the leading divergence when $p_1 \rightarrow 0$, and the behaviour is precisely the same as \eqref{soft123}. 

\subsection{Non-minimal form factors}

In this section, we verify in an explicit example that $n$-point   form factors of the operator $\cT_k$  (with $k<n$)   obey the same universal factorisation properties that hold for scattering amplitudes in general gauge theories 
\cite{Bern:1994zx,Bern:1993qk}, as well as for form factors of the stress tensor operator as shown in \cite{Brandhuber:2012vm}. 
This relation states that under the limit where two adjacent particles $a$ and $b$ with helicities $\sigma_a,\,\sigma_b$ become collinear, the $L$-loop $n$-point colour-ordered form factor (or amplitude) factorises into a sum of $(n-1)$-point form factors of equal or lower loop order, and the collinear divergences are encoded into the coefficients of each term -- the splitting amplitudes. For a general  form factor we have
\begin{align}
\label{eq:coll-general}
\begin{split}
\F_{\O,n}^{(L)}(1^{\sigma_1},\dots, a^{\sigma_a},b^{\sigma_b}, \dots, n^{\sigma_n}) 
\,{\buildrel a \parallel b\over
{\relbar\mskip-1mu\joinrel\longrightarrow}}
\, \sum_{\ell=0}^L\sum_{\sigma} \Big[  & \F^{(\ell)}_{\O,n-1} (1^{\sigma_1},\dots, (a+b)^{\sigma}, \dots, n^{\sigma_n})\\
& \times  \text{Split}_{-\sigma}^{(L-\ell)}(a^{\sigma_a},b^{\sigma_b}) \Big]\ , 
\end{split}
\end{align}
where $\sigma_i$ denote physical polarisations, and the sum is over all possible internal helicities~$\sigma$. 

\begin{figure}[h]
\centering
\includegraphics[width=0.4\textwidth]{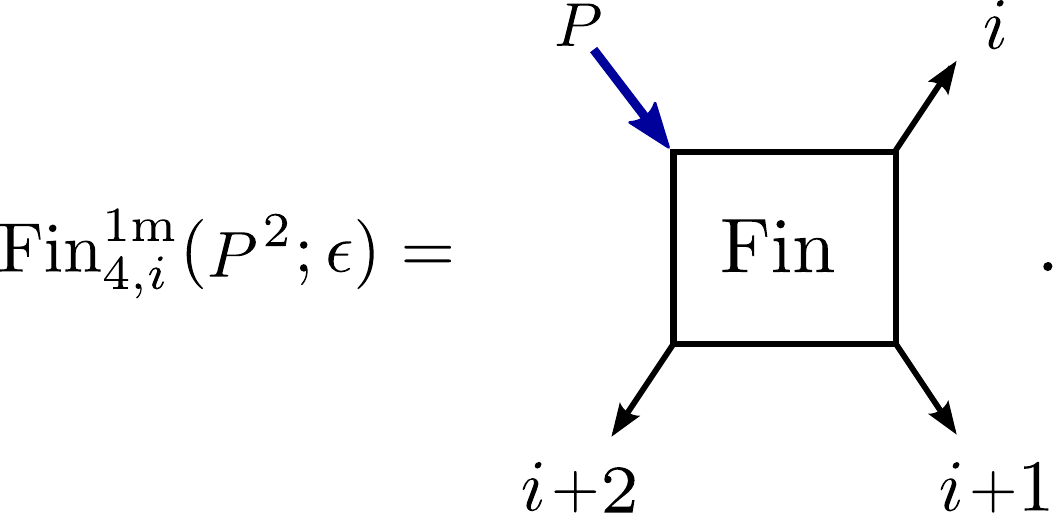}
\caption{\it One-mass finite box function with massless corner with momentum $P$.}
\label{fig:one-mass-fin-box}
\end{figure}

To confirm these factorisation properties, 
we will take as a representative example the particular component  form factor $F_{3,4}^{(1)}(1^+,2^{\phi_{12}},3^{\phi_{12}},4^{\phi_{12}};q)$, and will consider  the collinear limit $p_1\,||\,p_2$ defined in \eqref{collinear12}.  
For this case,  \eqref{eq:coll-general} predicts that 
\begin{align}
\label{eq:coll-one-loop}
\begin{split}
F_{3,4}^{(1)}(1^+,2^{\phi_{12}},3^{\phi_{12}},4^{\phi_{12}};q)\Big|_{1||2} \,&=\, F_{3,3}^{(0)}(P^{\phi_{12}},3^{\phi_{12}},4^{\phi_{12}};q)\,\text{Split}_{-\phi_{12}}^{(1)}(1^+,2^{\phi_{12}})\\
& +\,F_{3,3}^{(1)}(P^{\phi_{12}},3^{\phi_{12}},4^{\phi_{12}};q)\,\text{Split}_{-\phi_{12}}^{(0)}(1^+,2^{\phi_{12}})\ ,
\end{split}
\end{align}
where the tree-level and one-loop splitting functions with the helicities specified above are given by
\begin{align}
\label{eq:split-tree}
\text{Split}_{-\phi_{12}}^{(0)}(1^+,2^{\phi_{12}}) \, & = \,  \frac{1}{\b{12}} \sqrt{\frac{1-z}{z}}\ ,\\
\label{eq:split-1loop}
\text{Split}_{-\phi_{12}}^{(1)}(1^+,2^{\phi_{12}}) \, & = \, \frac{c_{\Gamma}}{\epsilon^2} (-s_{12})^{-\epsilon} \Big[1 - F\Big(\frac{z-1}{z}\Big) - F\Big(\frac{z}{z-1}\Big) \Big]\frac{1}{\b{12}} \sqrt{\frac{1-z}{z}}\ ,
\end{align}
and where we have introduced the shorthand notation $F(x)\,:=\,{_{2\!}F_1}(1,-\epsilon,1-\epsilon;x)$. The form factors appearing on the right-hand side of \eqref{eq:coll-one-loop} are given by
\begin{align}
\label{eq:RHS-FFs}
\begin{split}
F_{3,3}^{(0)}(P^{\phi_{12}},3^{\phi_{12}},4^{\phi_{12}};q) \,=\,& 1\ , \\[6pt]
F_{3,3}^{(0)}(P^{\phi_{12}},3^{\phi_{12}},4^{\phi_{12}};q) \,=\,&  -\frac{c_{\Gamma}}{\epsilon^2} \big[(-s_{P3})^{-\epsilon}+(-s_{34})^{-\epsilon}+(-s_{4P})^{-\epsilon}\big] \ .
\end{split}
\end{align}
In order to check \eqref{eq:coll-one-loop}, we use the general expression for the super form factors of $\T_3$ given in \eqref{eq:1-loop-MHV-BIS}. For the case of $F_{3,3}^{(1)}(1^+,2^{\phi_{12}},3^{\phi_{12}},4^{\phi_{12}};q)$, \eqref{eq:1-loop-MHV-BIS} reduces to
\begin{align}
\label{eq:F41loop}
\begin{split}
F^{(1)}_{3,4}(1^+,2^{\phi_{12}},3^{\phi_{12}},4^{\phi_{12}};q) =\  -\frac{c_{\Gamma}}{\epsilon^2} \frac{\b{24}}{\b{12}\b{14}} \big[(-s_{12})^{-\epsilon}+(-s_{23})^{-\epsilon}+(-s_{34})^{-\epsilon}+(-s_{41})^{-\epsilon}\big]& \\
 + \frac{\b{34}}{\b{31}\b{41}}\text{Fin}^{1{\rm m}}_{4,3}(s_{341};\epsilon)+ \frac{\b{24}}{\b{12}\b{14}}\text{Fin}^{1{\rm m}}_{4,4}(s_{412};\epsilon)+ \frac{\b{23}}{\b{13}\b{12}}\text{Fin}^{1{\rm m}}_{4,2}(s_{234};\epsilon)&\ ,
\end{split}
\end{align}
where $\text{Fin}_{4,i}^{1{\rm m}}(P^2;\epsilon)$ stands for the one-mass finite box function shown in Figure \ref{fig:one-mass-fin-box}, and is given by
\begin{align}
\begin{split}
&\text{Fin}_{4,i}^{1{\rm m}}(P^2;\epsilon)\,=\,-\frac{c_\Gamma}{\epsilon^2}\big[ (-s)^{-\epsilon}h(a\,s) + (-t)^{-\epsilon}h(a\,t)-(-P^2)^{-\epsilon} h(a\,P^2)\big]\ ,\\[5pt]
&s=s_{i\,i+1},\quad t=s_{i+1\,i+2},\quad u=s_{i\,i+2},\quad P^2=s_{i\,i+1\,i+2},\quad a\,:=\,-\frac{u}{st}\ ,
\end{split}
\end{align}
where we have defined  $ h(x)\, :=\, {_{2\!}F_1} \left(1,-\epsilon,1-\epsilon,x\right)-1 $ . 
\\
Under the collinear limit $p_1\,||\,p_2$, we first notice that the term $\dfrac{\b{34}}{\b{31}\b{41}}\text{Fin}^{1{\rm m}}_{4,3}(s_{341};\epsilon) $ in \eqref{eq:F41loop} is subleading,  and the remaining terms give
\begin{align}
\label{eq:F4coll}
\begin{split}
& F_{3,4}^{(1)}(1^+,2^{\phi_{12}},3^{\phi_{12}},4^{\phi_{12}};q)\Big|_{1||2}\, =\   -\frac{c_{\Gamma}}{\epsilon^2} \sqrt{\frac{1-z}{z}}\frac{1}{\b{12}}  \Big[ (-s_{12})^{-\epsilon}+(-s_{3P}(1-z))^{-\epsilon}+(-s_{34})^{-\epsilon}\\
& +(-s_{4P}z)^{-\epsilon} +(-s_{4P}z)^{-\epsilon}h\Big((z-1)\frac{s_{4P}}{s_{12}}\Big)+ (-s_{12})^{-\epsilon}h\Big(\frac{z-1}{z}\Big)-(-s_{4P})^{-\epsilon}h\Big(\frac{z-1}{z}\frac{s_{4P}}{s_{12}}\Big)\\
&+ (-s_{12})^{-\epsilon}h\Big(\frac{z}{z-1}\Big)+ (-s_{3P}(1-z))^{-\epsilon}h\Big(-z\frac{s_{3P}}{s_{12}}\Big)-(-s_{3P})^{-\epsilon}h\Big(\frac{z}{z-1}\frac{s_{3P}}{s_{12}}\Big)\Big]\ .
\end{split}
\end{align}
From \eqref{eq:F4coll} we already see the tree-level splitting amplitude \eqref{eq:split-tree} appearing as an overall prefactor. The terms with $(-s_{12})^{-\epsilon}$ combine to give the one-loop splitting amplitude \eqref{eq:split-1loop} as expected, 
\begin{align}
\begin{split}
F_{3,4}^{(1)}(1^+,2^{\phi_{12}},3^{\phi_{12}},4^{\phi_{12}};q)\Big|_{1||2}^{(-s_{12})^{-\epsilon}}\, =\  &\,\frac{c_{\Gamma}}{\epsilon^2} (-s_{12})^{-\epsilon} \sqrt{\frac{1-z}{z}}\frac{1}{\b{12}} \Big[1- F\Big(\frac{z-1}{z}\Big) - F\Big(\frac{z}{z-1}\Big) \Big]\\[6pt]
=&\ \text{Split}^{(1)}_{-{\phi_{12}}}(1^{\phi_{12}},2^{\phi_{12}})\,F^{(0)}_{3,3}(P^{\phi_{12}},3^{\phi_{12}},4^{\phi_{12}};q)\ ,\\[8pt]
\end{split}
\end{align}
whereas performing an expansion in $\epsilon$  of the remaining terms shows that it matches precisely $\,F_{3,3}^{(1)}(P^{\phi_{12}},3^{\phi_{12}},4^{\phi_{12}};q)\,\text{Split}_{-\phi_{12}}^{(0)}(1^+,2^{\phi_{12}})$. Thus we conclude that the universal collinear factorisation structure \eqref{eq:coll-general} is obeyed for the particular  one-loop form factor  we considered. Confirming the collinear factorisation at two-loop order would require the
calculation of the non-minimal form factor $F^{(2)}_{3,4}$,  which we leave for future investigations.

\vspace{0.3cm}

\section*{Acknowledgements}

It is a pleasure to thank  Bill Spence for initial collaboration and helpful discussions, and
Lance Dixon and Anastasia Volovich for interesting discussions. This work was supported by the Science and Technology Facilities Council Consolidated Grant ST/J000469/1  {\it String theory, gauge theory \& duality. }


\bibliographystyle{utphys}
\bibliography{bibliography}


\end{document}